\documentclass[journal=ancac3,manuscript=article]{achemso}

\usepackage[T1]{fontenc} 
\usepackage[version=3]{mhchem} 

\usepackage[utf8]{inputenc}
\usepackage[english]{babel}
\usepackage{graphicx}
\usepackage{dcolumn}
\usepackage{epsfig}
\usepackage{bm}
\usepackage{color}
\usepackage{siunitx}
\usepackage{amsmath}
\usepackage{enumerate} 
\usepackage{multirow}
\usepackage{upgreek}
\usepackage{longtable}
\usepackage{comment}
\usepackage{rotating}
\usepackage{array}

\author{Yuting Liu}
\affiliation{Department of Electronic and Computer Engineering, Hong Kong University of Science and Technology, Clear Water Bay, Kowloon, Hong Kong, China}
\author{Qiming Shao}
\affiliation{Department of Electronic and Computer Engineering, Hong Kong University of Science and Technology, Clear Water Bay, Kowloon, Hong Kong, China}
\email{eeqshao@ust.hk}

\title[An \textsf{achemso} demo]
  {Two-Dimensional Materials for Energy-Efficient Spin-Orbit Torque Devices}

\begin{document}






\begin{abstract}
Spin-orbit torques (SOTs), which rely on spin current generation from charge current in a nonmagnetic material, promise an energy-efficient scheme for manipulating magnetization in magnetic devices.  A critical topic for spintronic devices using SOTs is to enhance the charge to spin conversion efficiency. Besides, the current-induced spin polarization is usually limited to in-plane, whereas out-of-plane spin polarization could be favored for efficient perpendicular magnetization switching. Recent advances in utilizing two important classes of two-dimensional materials$-$topological insulators and transition-metal dichalcogenides$-$as spin sources to generate SOT shed light on addressing these challenges. Topological insulators such as bismuth selenide have shown a giant SOT efficiency, which is larger than those from three-dimensional heavy metals by at least one order of magnitude. Transition-metal dichalcogenides such as tungsten telluride have shown a current-induced out-of-plane spin polarization, which is allowed by the reduced symmetry. In this review, we use symmetry arguments to predict and analyze SOTs in two-dimensional material-based heterostructures. We summarize the recent progress of SOT studies based on topological insulators and transition-metal dichalcogenides and show how these results are in line with the symmetry arguments. At last, we identify unsolved issues in the current studies and suggest three potential research directions in this field.

\end{abstract}


Ubiquitous smart devices and the Internet of Things create tremendous data every day, shifting computing diagram towards data-driven. Computing and memory units in traditional computers are physically separate, which leads to huge energy cost and time delay. Emerging computer architectures bring computing and memory units together for data-intensive applications. Magneto-resistive random-access memory (MRAM) is the leading contender for future embedded nonvolatile memory due to its high speed, energy efficiency, and theoretically unlimited endurance.\cite{kent_new_2015} MRAM and many other magnetic or spintronic devices rely on the efficient manipulation of magnetic moments through electrically generated spin current. Looking for efficient methods to generate spin current with the least possible charge current is an essential task for spintronics. Recently, in nonmagnetic materials with strong spin-orbit coupling, spin Hall effect or Rashba effect has been shown to generate a large spin current and thus a strong spin-orbit torque (SOT) on the adjacent ferromagnet (FM), enabling magnetic domain wall motion, magnetization switching, and magnetic resonance.\cite{sinova_spin_2015,miron_perpendicular_2011,liu_spin-torque_2012,manchon_current-induced_2019} As a result, SOT-based devices, such as SOT-MRAM, magnetic nonvolatile logic,\cite{wang_electric-field_2016} and nano-oscillators\cite{chen_spin-torque_2016} have attracted widespread attention for ultralow-power nonvolatile memory and logic applications. \\
Commonly used SOT source materials are three-dimensional (3D) heavy metals (HM), such as Pt, Ta, and W, which face two challenges. $First$, in most cases, the damping-like SOT plays a dominant role in magnetization switching since it compensates the magnetic damping. The damping-like SOT efficiencies of HM range from 0.1 to 0.3.\cite{manchon_current-induced_2019} To reduce the switching energy, improved SOT efficiency is required. $Second$, in HM/FM heterostructures with perpendicular magnetic anisotropy (PMA), the damping-like SOT generated by HMs lies in the film plane as required by a global twofold rotational symmetry, which does not allow a deterministic and energy-efficient switching without an external magnetic field.\cite{yu_switching_2014} For MRAM technology, perpendicularly magnetized magnets are preferred because they require a much lower switching current compared with that of in-plane magnetized magnets. To achieve deterministic and energy-efficient switching of a magnet with PMA, the out-of-plane damping-like SOT is desired.  \\
Recently, two important classes of two-dimensional (2D) materials emerged for generating SOTs, which allow one to address the above-mentioned two challenges. The first class is the 2D Bi$_2$Se$_3$ family, which are 3D topological insulators (TIs). TIs exhibit a giant damping-like SOT efficiency larger than 100 at low temperatures due to the spin momentum locking of topological surface states.\cite{fan_magnetization_2014} The orders of magnitude of improvement of SOT efficiency can potentially bring a much lower switching current density for SOT devices. The second class is 2D transition-metal dichalcogenides (TMDs), some of which have a very strong spin-orbit coupling and much lower global symmetries than 3D HMs. For example, a low-symmetry 2D TMD, T$_\textnormal{d}$-WTe$_2$ with an orthorhombic lattice structure allows for the  generation of out-of-plane damping-like SOT,\cite{macneill_control_2017} which could potentially enable deterministic and energy-efficient switching of the perpendicular magnet. \\
2D materials offer many exceptional properties for spintronic applications, including but not limited to electrical\cite{fan_electric-field_2016} and chemical\cite{balakrishnan_colossal_2013} tunability of spintronic properties, intrinsic ferromagnetism\cite{gong_discovery_2017,huang_layer-dependent_2017} and antiferromagnetism,\cite{huang_layer-dependent_2017} long spin diffusion length,\cite{han_graphene_2014,song_coexistence_2020} large and multidirectional spin-to-charge conversion efficiency,\cite{safeer_large_2019} and large charge-to-spin conversion or SOT efficiency.\cite{fan_magnetization_2014,mellnik_spin-transfer_2014} In this review, we highlight two key advantages regarding SOTs. First, 2D materials are crystalline materials down to the monolayer limit and they have diverse band structures. Many 2D materials with a strong spin-orbit coupling show topological properties, which are appealing to generating large SOTs. For example, T$_\textnormal{d}$-WTe$_2$ is Weyl semimetal in the bulk form\cite{soluyanov_type-ii_2015} and 2D TI in the monolayer limit.\cite{qian_quantum_2014} Second, 2D materials have diverse symmetry properties, which could be used to guide researchers to identify unconventional SOTs, $i.e.$, SOTs with unconventional directions. For example, while an out-of-plane damping-like SOT cannot be generated by 2H-TMDs that have preserved twofold rotational symmetry, it can be generated by 1T’- or T$_\textnormal{d}$-TMDs that have a broken twofold rotationary symmetry. Therefore, studying SOTs generated by 2D materials is of critical importance for spintronic applications. \\
Motivated by the breakthroughs in SOT generated by TIs and TMDs, intensive investigations of SOT devices based on 2D materials are ongoing. SOTs have been experimentally studied in heterostructures composed of an FM and a 2D material, including TIs (Bi$_2$Se$_3$ family,\cite{fan_magnetization_2014,fan_electric-field_2016,mellnik_spin-transfer_2014}) semiconducting TMDs (2H-MoS$_2$\cite{zhang_research_2016,shao_strong_2016} and 2H-WSe$_2$,\cite{shao_strong_2016}) metallic TMDs (T$_\textnormal{d}$-WTe$_2$,\cite{macneill_control_2017,li_spin-momentum_2018} 1T’-MoTe$_2$,\cite{stiehl_layer-dependent_2019} 1T’-TaTe$_2$,\cite{stiehl_current-induced_2019}, PtTe$_2$\cite{xu_high_2020}) and superconducting TMD (NbSe$_2$\cite{guimaraes_spin_2018}). While the number of experimental reports about SOTs generated by 2D materials has dramatically increased, there is no systematic review of symmetry analysis and experimental results that could help identify the key open questions and guide further studies.\\ 
In this review, we first present the SOT prediction based on symmetry argument and show the theoretical predictions of SOTs generated by various 2D materials according to their symmetry properties. Then, we review recent experimental progress about SOT studies in 2D material-based heterostructures. We will focus on TI and TMD systems. Through this summary, we will identify how the symmetry analysis aligns with the experimental observation and what is missing in current studies. Besides, we review the current understanding of the origins of SOTs in these heterostructures and the progress of large-scale growth of high-quality 2D materials for spintronic applications. At last, we provide a conclusion and outlook that briefly summarizes the article and points out three potential research directions.

\section{Background of Research}
\begin{figure}[ht]
	\centering
	\includegraphics[width=16cm]{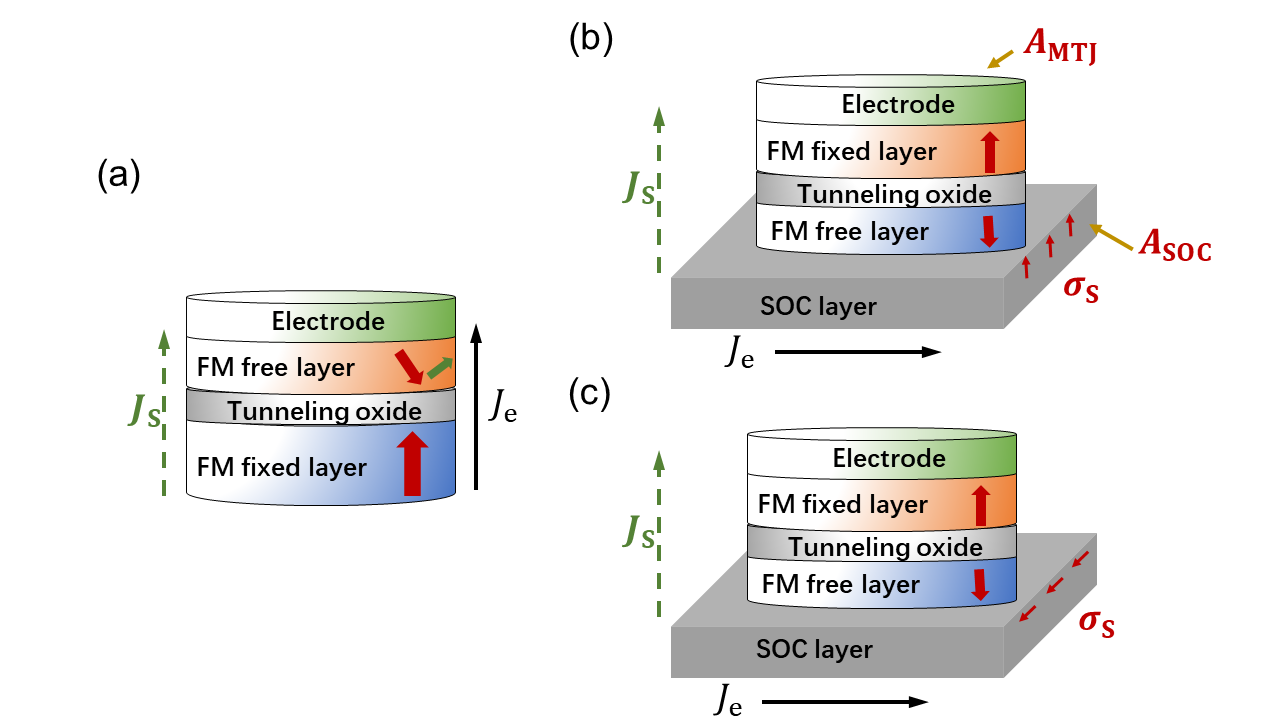}
	\vspace{0pt}
	\caption{Schematics of MRAMs. (a) STT-MRAM unit cell, (b)  and (c) SOT-MRAM unit cells with current-induced out-of-plane and in-plane spin polarization. Only the spin polarization near the top surface of the SOC layer has been shown in (b) and (c). }
	\label{figMRAM}
\end{figure}

There are two important schemes for generating spin torques: spin-transfer torques (STTs) and SOTs. Here, we will use a comparison between these two schemes to show the significance of a large out-of-plane damping-like torque for switching perpendicular magnetization. STTs are originated from spins polarized by a magnetic material, which can be transferred to another magnetic material to exert a spin torque. The current-induced spin polarization unit vector $\bm{\sigma}$ is collinear with the magnetization unit vector ($\bm{m}$) of the ferromagnetic fixed layer. STTs were first theoretically proposed by Slonczewski\cite{slonczewski_current-driven_1996} and Berger\cite{berger_emission_1996} in 1996 and then later realized experimentally in 1999.\cite{katine_current-driven_2000,myers_current-induced_1999} There are two types of STTs: field-like (FL) or adiabatic STT and Slonczewski or non-adiabatic (anti-)damping-like (DL) STT. FL-STT and DL-STT are written as $\bm{\vec{\uptau}_{\textnormal{FL}}}$ = $\uptau_{\textnormal{FL}} \bm{m} \times \bm{\sigma}$ and $\bm{\vec{\uptau}_{\textnormal{DL}}}$ = $\uptau_{\textnormal{DL}} \bm{m} \times (\bm{m} \times \bm{\sigma)}$, respectively, where $\uptau_{\textnormal{FL}}$ and $\uptau_{\textnormal{DL}}$ are the magnitude of FL-STT and DL-STT, respectively. In the equilibrium, $\bm{m}$'s of the fixed layer and the free layer are collinear (parallel or antiparallel), resulting in zero STTs. Thus, STT driven switching is normally initiated by thermal effects which limit STT switching time and minimum error rate. \cite{yu_two-terminal_2018}\\
SOTs mainly result from spin-orbit coupling (SOC) of a nonmagnetic material, which allows a charge current to be converted to a spin current and then exert a spin torque on the adjacent magnetic material. FL-SOT and DL-SOT have the same expressions as the FL-STT and DL-STT. However, the current-induced spin polarization \bm{$\sigma$} in the SOT case is originated from the SOC and thus its direction is not necessarily collinear with the initial \bm{$m$} of the ferromagnetic free layer. While there were early theoretical studies about SOTs in 2008-2009\cite{manchon_theory_2008,manchon_theory_2009} and experimental demonstration of SOT-induced magnetization re-orientation in (Ga,Mn)As in 2009\cite{chernyshov_evidence_2009} potential and relevance of SOTs for real applications are brought to attention by room-temperature SOT-induced, magnetization switching in Pt/Co/AlO$_x$ in 2011\cite{miron_perpendicular_2011} and Ta/CoFeB/MgO-based magnetic tunnel junctions (\textnormal{MTJ}s) in 2012.\cite{liu_spin-torque_2012}\\
There are two important advantages of SOTs in comparison to STTs. First, for STT-MARM (see Fig. \ref{figMRAM} (a)), the writing current needs to go through the \textnormal{MTJ}, causing the easy breakdown of nanometer-thick MgO tunnel barrier in the \textnormal{MTJ}. In contrast, for SOT-MRAM (Fig. \ref{figMRAM} (b)), the writing current is passing through the metal line instead of \textnormal{MTJ}s, greatly improving the endurance of the SOT-MRAM. Second, STT efficiency ($\eta$) is fundamentally limited by the spin polarization at the Fermi level of ferromagnetic metal, which cannot be larger than one. However, recent advances have shown that SOT efficiency ($\xi$) can be larger than one in some topological materials, such as 3D topological insulators Bi$_2$Se$_3$ \cite{mellnik_spin-transfer_2014} and (BiSb)$_2$Te$_3$.\cite{fan_magnetization_2014} A comprehensive introduction to STT- and SOT-MRAM can be found in Wang $et$ $al$..\cite{wang_low-power_2013,wang_electric-field_2016} \\

The comparison between critical currents of SOT- and STT-MRAM ($I_{\textnormal{SOT}}$ and $I_{\textnormal{STT}}$) under the assumption of a single-domain magnet can be written as \cite{wang_low-power_2013}
\begin{equation} \label{EQx}
\frac{I_{\textnormal{SOT}}}{I_{\textnormal{STT}}} \approx \frac{\eta}{\xi} \frac{A_{\textnormal{SOC}}}{A_{\textnormal{MTJ}}} ,
\end{equation}
where $A_{\textnormal{SOC}}$ and $A_{\textnormal{MTJ}}$ are the cross-section area of the SOC layer and \textnormal{MTJ} layer, respectively. As shown in Fig. \ref{figMRAM} (b), the width of the SOC layer will be at the order of \textnormal{MTJ} diameter and the thickness of the SOC layer is typically several nanometers, which is much smaller than typical \textnormal{MTJ} diameter (tens to hundreds of nanometers). Therefore,  $A_{\textnormal{SOC}}/A_{\textnormal{MTJ}} <1$. From Eq. \ref{EQx}, if $\eta/\xi<1$, we will have $I_{\textnormal{SOT}}/I_{\textnormal{STT}} <1$. This suggests that the SOT-MRAM has a significant advantage regarding the writing current over the STT-MRAM if one can utilize a SOC layer with a large $\xi$. However, Eq. \ref{EQx} has an assumption that is not valid in most of the SOT experiments. The assumption is that the current-induced spin polarization is collinear with the initial magnetization of the free layer, which is always true for STT-MRAM but not for SOT-MRAM. To date, most of the thin film SOC materials, such as Pt, Ta, and W, can only generate a spin polarization lying in the film plane, which is orthogonal to the initial $\bm{m}$ of a free layer with PMA as shown in Fig. \ref{figMRAM} (c). Note that the ferromagnetic layers with PMA are preferred for MRAM due to a much better scalability and lower writing energy.\cite{wang_low-power_2013} In these cases, the ratio of $I_{\textnormal{SOT}}$ to $I_{\textnormal{STT}}$ (single-domain magnet consumption) is given by \cite{lee_threshold_2013,macneill_control_2017,liu_current-induced_2012}
\begin{equation}\label{EQx1}
\frac{I_{\textnormal{SOT}}}{I_{\textnormal{STT}}} \approx \frac{\eta}{2\alpha \xi} \frac{A_{\textnormal{SOC}}}{A_{\textnormal{MTJ}}} ,
\end{equation}
where $\alpha$ is the Gilbert damping factor. Since $\alpha$ is typically on the order of 0.01, $\xi$ needs to be orders of magnitude larger than $\eta$ to achieve a smaller $ I_{\textnormal{SOT}}$ compared with $ I_{\textnormal{STT}}$. So we need to figure out how to generate an out-of-plane spin polarization in SOT-MRAM to gain the energy advantage shown in Eq. \ref{EQx}. Moreover, the thermal incubation time is not needed for SOT switching when both the in-plane and out-of-plane spin polarizations exist.  In the following, we will use symmetry arguments to understand why there is only an in-plane spin polarization in HM/FM heterostructures and how we can generate an out-of-plane spin polarization.\\
In a material or material heterostructure with a sizable SOC, a spin polarization can be induced when the inversion symmetry is broken. Normally, sputtered HMs like Pt and Ta have a global inversion symmetry despite the fact that there is no microscopic symmetry due to the amorphous or polycrystalline nature of these materials. Therefore, they cannot allow for a net spin polarization in the bulk form. However, when an HM is interfacing to an FM like Co or CoFeB, the inversion symmetry is broken and at the interface, spin polarizations can be induced by applying an electric current in the interface plane. This effect can also be understood using a Rashba picture: at the interface, there is a vertical electric field \bm{$E_z$} along the out-of-film-plane direction; when an in-plane electric current is applied, there is an induced spin accumulation $\bm{\sigma} \propto \bm{j} \times \bm{E_z}$,\cite{manchon_new_2015} which is orthogonal to both the out-of-plane direction and the current direction. While we here describe the effect of current-induced spin polarization using a Rashba picture, in reality, the effect can be due to spin Hall effect, Rashba effect, and Dresselhaus effect, where the latter two are also called inverse spin galvanic effect. \cite{manchon_current-induced_2019} Note that since there is a rotation symmetry in the film plane, the current-induced out-of-plane spin polarization is not allowed as we will describe below.\\

\section{Crystal Symmetry and SOT}

\begin{table}
    
    \centering
    \small
		\begin{tabular}{p{1.5cm}cp{1.5cm}c}
			\hline \hline
			Point Group&$\chi$&Point Group&$\chi$\\ \hline
			C$_1$&
			 $\begin{pmatrix}
		     \chi_{\rm{xx}}&\chi_{\rm{xy}}&\chi_{\rm{xz}}\\
	 	     \chi_{\rm{yx}}&\chi_{\rm{yy}}&\chi_{\rm{yz}}\\
		     \chi_{\rm{zx}}&\chi_{\rm{zy}}&\chi_{\rm{zz}}\\
             \end{pmatrix}$&
             C$_2$&
             $\begin{pmatrix}
		     \chi_{\rm{xx}}&\chi_{\rm{xy}}&0\\
	 	     \chi_{\rm{yx}}&\chi_{\rm{yy}}&0\\
		     0&0&\chi_{\rm{zz}}\\
             \end{pmatrix}$\\
			 C$_3$&
            $\begin{pmatrix}
		     \chi_{\rm{xx}}&\chi_{\rm{xy}}&0\\
	 	    \chi_{\rm{yx}}[=-\chi_{\rm{xy}}]&\chi_{\rm{yy}}[=\chi_{\rm{xx}}]&0\\
	        0&0&\chi_{\rm{zz}}\\
            \end{pmatrix}$&
			 C$_4$&
            $\begin{pmatrix}
		     \chi_{\rm{xx}}&\chi_{\rm{xy}}&0\\
	 	    \chi_{\rm{yx}}[=-\chi_{\rm{xy}}]&\chi_{\rm{yy}}[=\chi_{\rm{xx}}]&0\\
	        0&0&\chi_{\rm{zz}}\\
            \end{pmatrix}$\\
			 C$_6$&
			 $\begin{pmatrix}
		     \chi_{\rm{xx}}&\chi_{\rm{xy}}&0\\
	 	     \chi_{\rm{yx}}[=-\chi_{\rm{xy}}]&\chi_{\rm{yy}}[=\chi_{\rm{xx}}]&0\\
		     0&0&\chi_{\rm{zz}}\\
            \end{pmatrix}$&
			 C$_{\rm{S}}$&
			 $\begin{pmatrix}
		     0&\chi_{\rm{xy}}&0\\
	 	     \chi_{\rm{yx}}&0&\chi_{\rm{yz}}\\
		     0&\chi_{\rm{zy}}&0\\
             \end{pmatrix}$\\
			 C$_{2v}$&
			 $\begin{pmatrix}
		     0&\chi_{\rm{xy}}&0\\
	 	     \chi_{\rm{yx}}&0&0\\
		     0&0&0\\
             \end{pmatrix}$&
			 C$_{3v}$&
			 $\begin{pmatrix}
		     0&\chi_{\rm{xy}}&0\\
	 	     \chi_{\rm{yx}}[=-\chi_{\rm{xy}}]&0&0\\
		     0&0&0\\
		     \end{pmatrix}$\\
		     C$_{4v}$&
			 $\begin{pmatrix}
		     0&\chi_{\rm{xy}}&0\\
	 	     \chi_{\rm{yx}}[=-\chi_{\rm{xy}}]&0&0\\
		     0&0&0\\
		     \end{pmatrix}$&
		     C$_{6v}$&
			 $\begin{pmatrix}
		     0&\chi_{\rm{xy}}&0\\
	 	     \chi_{\rm{yx}}[=-\chi_{\rm{xy}}]&0&0\\
		     0&0&0\\
		     \end{pmatrix}$\\
		     D$_{2d}$&
			 $\begin{pmatrix}
		     \chi_{\rm{xx}}&0&0\\
	 	     0&\chi_{\rm{yy}}[=-\chi_{\rm{xx}}]&0\\
		     0&0&0\\
		     \end{pmatrix}$&		     
		     S$_{4}$&
			 $\begin{pmatrix}
		     \chi_{\rm{xx}}&\chi_{\rm{xy}}&0\\
	 	     \chi_{\rm{yx}}[=\chi_{\rm{xy}}]&\chi_{\rm{yy}}[=-\chi_{\rm{xx}}]&0\\
		     0&0&0\\
		     \end{pmatrix}$\\
		     D$_{2}$&
			 $\begin{pmatrix}
		     \chi_{\rm{xx}}&0&0\\
	 	     0&\chi_{\rm{yy}}&0\\
		     0&0&\chi_{\rm{zz}}\\
		     \end{pmatrix}$&
		     D$_{3}$&
			 $\begin{pmatrix}
		     \chi_{\rm{xx}}&0&0\\
	 	     0&\chi_{\rm{yy}}[=\chi_{\rm{xx}}]&0\\
		     0&0&\chi_{\rm{zz}}\\
		     \end{pmatrix}$\\
		     D$_{4}$&
			 $\begin{pmatrix}
		     \chi_{\rm{xx}}&0&0\\
	 	     0&\chi_{\rm{yy}}[=\chi_{\rm{xx}}]&0\\
		     0&0&\chi_{\rm{zz}}\\
		     \end{pmatrix}$&
		     D$_{6}$&
			 $\begin{pmatrix}
		     \chi_{\rm{xx}}&0&0\\
	 	     0&\chi_{\rm{yy}}[=\chi_{\rm{xx}}]&0\\
		     0&0&\chi_{\rm{zz}}\\
		     \end{pmatrix}$\\
		     T&
			 $\begin{pmatrix}
		     \chi_{\rm{xx}}&0&0\\
	 	     0&\chi_{\rm{yy}}[=\chi_{\rm{xx}}]&0\\
		     0&0&\chi_{\rm{zz}}[=\chi_{\rm{xx}}]\\
		     \end{pmatrix}$&
		     O&
		     $\begin{pmatrix}
		     \chi_{\rm{xx}}&0&0\\
	 	     0&\chi_{\rm{yy}}[=\chi_{\rm{xx}}]&0\\
		     0&0&\chi_{\rm{zz}}[=\chi_{\rm{xx}}]\\
		     \end{pmatrix}$\\
		     
			\hline\hline
		\end{tabular}
	    \caption{ Magnetoelectric pseudotensor $\chi$ in spin-orbit coupled material/globally high-symmetry FM thin film heterostructures with different group symmetries.\cite{he_novel_2020} The principle axis is along $z$ axis. }
		\label{tab_pointgroup}
		\end{table}


It is essential to determine the impact of crystal symmetry on the SOTs of 2D/FM materials for the SOT prediction. Since DL- and FL-SOTs ($\bm{\vec{\uptau}_{\textnormal{DL}}}$ and $\bm{\vec{\uptau}_{\textnormal{FL}}}$) are directly related to $\bm{\sigma}$ by following $\uptau_{\textnormal{DL}} \bm{m} \times (\bm{m} \times \bm{\sigma)}$ and $\uptau_{\textnormal{FL}} \bm{m} \times \bm{\sigma}$, respectively,  we will determine the current-induced spin polarization using symmetry arguments. Conventional SOTs refer to SOTs resulting from the in-plane spin polarization, while the unconventional SOTs refer to SOTs resulting from the out-of-plane spin polarization. We adopt two methods, one is referred to as symmetry operation analysis which was used by MacNeil $et$ $al.$ \cite{macneill_control_2017} and the other one is referred to as symmetry matrix analysis which was used by He and Law,\cite{he_novel_2020} for determining the current-induced spin polarization using symmetry arguments. \\
\subsection{Inversion Symmetry}
We mentioned above that current-induced spin polarizations cannot exist when the inversion symmetry is preserved. This can be understood using the symmetry operation analysis: the spin polarization is a pseudovector that does not change sign and electric field (current) is a vector that changes sign under an inversion operation. In other words, the inversion symmetry leads to $\bm{\sigma} \rightarrow \bm{\sigma}$ and $\bm{j} \rightarrow -\bm{j}$, where \bm{$j$} is the applied electric current. Since the induced spin polarization is proportional to current to the first order ($\bm{\sigma} \propto \bm{j}$), $\bm{\sigma} \propto -\bm{j}$ under inversion. Therefore, $\bm{\sigma}$=0 to the linear order. The second way to understand this is by using general symmetry analysis. The induced spin polarization vector ($\bm{\sigma}$) will be  $\bm{\sigma} =\chi \bm{j}$, where $\chi$ is the 3D magnetoelectric pseudotensor. If a system has a set of symmetry operations $\{R\}$, then $\chi$ should respect
\begin{equation} \label{EQsym1}
\chi = {\rm{det}} (R_i)R_i \chi R_i^T,R_i \in \{R\}
\end{equation}
where $R_i$ is the arbitrary symmetry operation in the set, det$(R_i)$ is its determinant, and $R_i^T$ is its transpose ($R_i R_i^T$=1).  For readers who are not familiar with symmetry groups, illustrations of symmetry groups can be found on reference\cite{noauthor_symmetryotterbein_nodate} and all symmetry operations of a point group can be easily found on reference.\cite{noauthor_character_nodate} To calculate the most general form of $\chi$, we first assume that
\begin{equation} \label{EQsym2}
 \chi  = \begin{pmatrix}
		\chi_{\rm{xx}}&\chi_{\rm{xy}}&\chi_{\rm{xz}}\\
		\chi_{\rm{yx}}&\chi_{\rm{yy}}&\chi_{\rm{yz}}\\
		\chi_{\rm{zx}}&\chi_{\rm{zy}}&\chi_{\rm{zz}}\\
\end{pmatrix}   
\end{equation}
where 9 elements in the tensor can be all arbitrary. We then calculate $\chi$ by  substituting Eq. \ref{EQsym2} and $R_i$ into Eq. \ref{EQsym1}. To validate the equation, some elements of $\chi$ will have to be zero and thus are not allowed by the symmetry operation $R_i$. As shown by He and Law,\cite{he_novel_2020} only 18 point groups allow for a non-zero spin accumulation in the presence of a charge current (see Table \ref{tab_pointgroup}). These 18 point groups are non-centrosymmetric since a net spin accumulation is not allowed in a system with inversion symmetry. Since the inversion symmetry operation matrix is given by
\begin{equation*}
    i=\begin{pmatrix}
    -1&0&0  \\
    0&-1&0\\
    0&0&-1
    \end{pmatrix}
\end{equation*}
Eq. \ref{EQsym2} under the inversion symmetry becomes:
\begin{equation}
\begin{split}
 \begin{pmatrix}
    \chi_{\textnormal{xx}}&\chi_{\rm{xy}}&\chi_{\rm{xz}}\\
    \chi_{\rm{yx}}&\chi_{\rm{yy}}&\chi_{\rm{yz}}\\
    \chi_{\rm{zx}}&\chi_{\rm{zy}}&\chi_{\rm{zz}} 
 \end{pmatrix}  = \rm{det}(\textit{i})\textit{i}\chi \textit{i}^T =-
  \begin{pmatrix}
    \chi_{\rm{xx}}&\chi_{\rm{xy}}&\chi_{\rm{xz}}\\
    \chi_{\rm{yx}}&\chi_{\rm{yy}}&\chi_{\rm{yz}}\\
    \chi_{\rm{zx}}&\chi_{\rm{zy}}&\chi_{\rm{zz}} 
 \end{pmatrix}
 \Longrightarrow
   \begin{pmatrix}
    \chi_{\rm{xx}}&\chi_{\rm{xy}}&\chi_{\rm{xz}}\\
    \chi_{\rm{yx}}&\chi_{\rm{yy}}&\chi_{\rm{yz}}\\
    \chi_{\rm{zx}}&\chi_{\rm{zy}}&\chi_{\rm{zz}} 
 \end{pmatrix} =0
 \end{split}
\end{equation}
Therefore, we see that the magnetoelectric tensor $\chi$ is zero for inversion symmetric systems. Note that for inversion symmetric systems, non-zero spin accumulations at the top and bottom surfaces with opposite signs can be induced in the presence of charge current. However, the net spin accumulation is zero as constrained by the inversion symmetry. In general, a spin Hall conductivity associated with the spin Hall effect is usually described by a rank-3 tensor\cite{seemann_symmetry-imposed_2015,wimmer_spin-orbit-induced_2015} and we need to reduce it to a rank-2 tensor by fixing the spin current direction to be out-of-plane direction if we want to continue the discussion as we do here. In the following, the impact of rotation, mirror, screw, and glide symmetries on SOTs will be discussed.\\

\subsection{Mirror Symmetry}
\begin{figure}
    \centering
    \includegraphics[width=16cm]{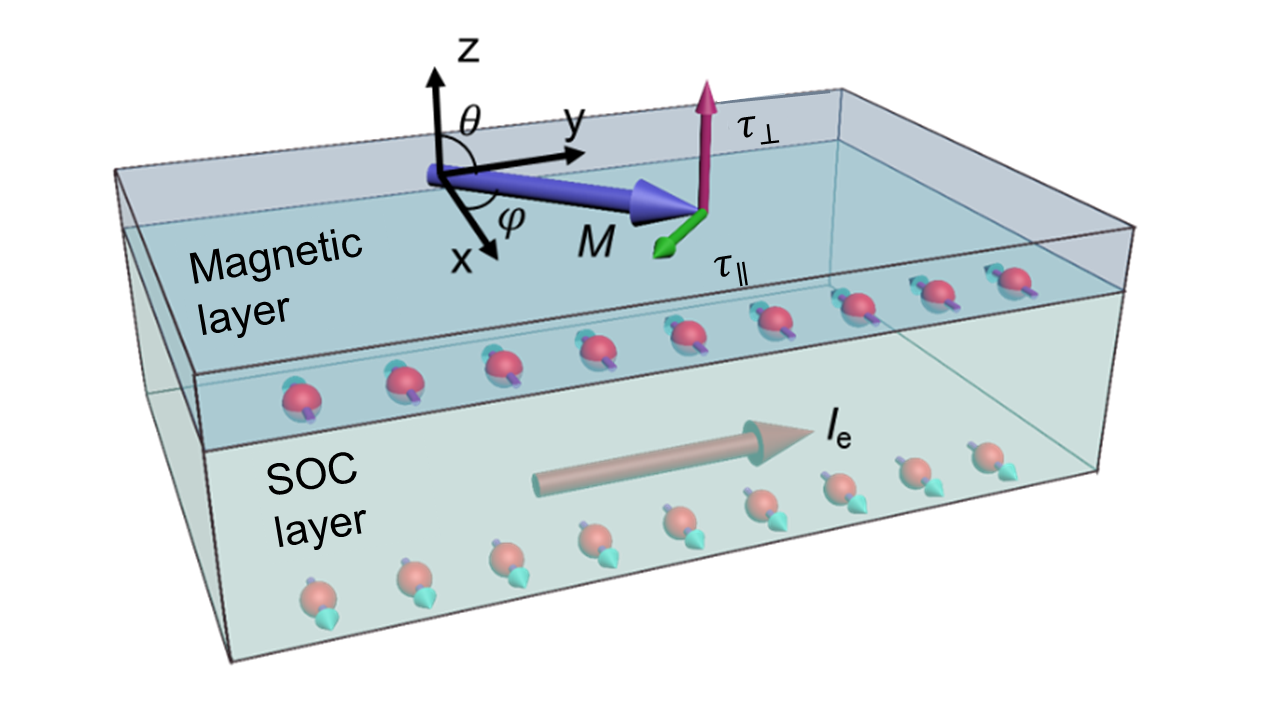}
    \caption{Schematic of a nonmagnetic spin-orbit coupled (SOC) layer/magnetic layer heterostructures for symmetry analysis of current-induced SOTs. Assuming that the current is flowing along the $y$ direction, the angle between magnetization and $x$ direction is azimuthal angle $\varphi$ and the angle between magnetization and $z$ direction is polar angle $\theta$. The red balls represent the electrons and the arrows represent the spin direction. Here, only conventional in-plane spin polarization and corresponding in-plane DL-SOT ($\uptau_{\parallel}$) and out-of-plane FL-SOT ($\uptau_{\perp}$) are illustrated.}
    \label{fig:SOCangle}
\end{figure}
The mirror plane perpendicular to the $z$ axis is broken in thin-film heterostructures.  Thus, we only consider the cases where there are mirror planes parallel to the $z$ axis of the heterostructure. When sending an in-plane current to the heterostructures, the simplified form of the torque can be expressed as:
	\begin{equation} \label{torque_form}
	\begin{split}
	\bm{\vec{\uptau}_{\perp}}(\bm{m},E) = \uptau_{\perp}(\varphi, E) \bm{z} \\
	\bm{\vec{\uptau}_{\parallel}}(\bm{m},E) = \uptau_{\parallel}(\varphi, E) \bm{m} \times \bm{z}, \\
	\end{split}
	\end{equation}
where $\bm{\vec{\uptau}_{\perp}}(\bm{m},E)$ and $\bm{\vec{\uptau}_{\parallel}(\bm{m},E)}$ are the out-of-plane and in-plane torque vectors, $E$ is the applied electric field, and $\varphi$ is the azimuth angle between the magnetization vector $\bm{m}$ and the $x$ axis as shown in Fig. \ref{fig:SOCangle}. Here, we assume that the magnetization is in the film plane. For cases where magnetization is out-of-plane, the symmetry operation analysis is similar except that the out-of-plane and in-plane torques are replaced by two orthogonal in-plane torques. The scalar-prefactor $\uptau_{\perp}(\varphi, E)$ and  $\uptau_{\parallel}(\varphi, E)$ can be Fourier expanded as \cite{macneill_control_2017}:
	\begin{equation}
	\begin{split} \label{torque_fourier}
	\uptau_{\perp} (\varphi, E) = E(A_0+A_1 \textnormal{sin}\varphi + A_2 \textnormal{cos}\varphi+ A_3 \textnormal{sin}2\varphi+A_4 \textnormal{cos}2\varphi+....) \\
	\uptau_{\parallel} (\varphi, E)= E(S_0+S_1 \textnormal{sin}\varphi+S_2 \textnormal{cos}\varphi+S_3 \textnormal{sin}2\varphi+S_4\textnormal{cos}2\varphi+....) ,\\
	\end{split}
	\end{equation}
where $A_1$ and $S_1$ represent the linear-order conventional FL- and DL-SOTs, respectively, due to the in-plane spin polarization and $A_0$ and $S_0$ represent the unconventional DL- and FL-SOTs, respectively, due to the out-of-plane spin polarization.  One can re-write Eq. \ref{torque_form} to obtain the symmetry transformation:
	\begin{equation} \label{torque_form2}
	\begin{split}
	\uptau_{\perp}(\varphi,E)= \bm{\vec{\uptau}_{\perp}} \cdot \bm{z} \\
	\uptau_{\parallel}(\varphi, E)= \bm{\vec{\uptau}_{\parallel}} \cdot (\bm{m} \times \bm{z}). \\
	\end{split}
	\end{equation}
For a lattice that preserves a mirror plane $M_{\textnormal{xz}}$ perpendicular to the $y$ axis, the mirror symmetry operation reverses the $x$ component of magnetization leading to $\varphi \rightarrow \pi-\varphi$ as shown in Fig. \ref{fig:SOCangle}. When sending an in-plane current along the $y$ direction, the electric field flips sign denoting as $E\rightarrow -E$. Since the dot and cross products of a pseudovector and a vector are pseudovector and vector, respectively, $\bm{\vec{\uptau}_{\perp}} \cdot \bm{z}$ transforms to $-\bm{\vec{\uptau}_{\perp}} \cdot \bm{z}$ and $\bm{\vec{\uptau}_{\parallel}} \cdot (\bm{m} \times \bm{z})$ transforms to $-\bm{\vec{\uptau}_{\parallel}}\cdot (\bm{m} \times \bm{z})$ under the mirror symmetry. Therefore, the mirror symmetry requires:
	\begin{equation} \label{torque_symmetry1}
	\begin{split}
	\uptau_{\perp}(\pi-\varphi, -E)= -\uptau_{\perp}(\varphi, E) \\
	\uptau_{\parallel}(\pi-\varphi, -E)= -\uptau_{\parallel}(\varphi, E) \\
	\end{split}
	\end{equation}
To fulfill the requirements of Eq. \ref{torque_symmetry1}, Eq. \ref{torque_form} is reduced to: 
	\begin{equation}
	\begin{split} \label{torque_ysymmetry}
	\uptau_{\perp} = E(A_0+A_1 \textnormal{sin}\varphi+A_3 \textnormal{sin}2\varphi+....), \\
	\uptau_{\parallel} = E(S_0+S_1 \textnormal{sin}\varphi+S_3 \textnormal{sin}2\varphi+....) .\\
	\end{split}
	\end{equation}
$A_0, A_1, S_0$ and $S_1$ are not reduced in the above equation which suggests that both conventional and unconventional torques are allowed if only $M_{\textnormal{xz}}$ exists and the current is along the y direction. \\
Now, if the lattice also preserves a mirror plane $M_{\textnormal{yz}}$ perpendicular to the $x$ axis, the symmetry transformation rules are: (1) $\varphi \rightarrow - \varphi$; (2) $E\rightarrow E$ and (3) $\bm{\vec{\uptau}_{\perp}} \cdot \bm{z} \rightarrow -\bm{\vec{\uptau}_{\perp}} \cdot \bm{z}$ and $\bm{\vec{\uptau}_{\parallel}} \cdot (\bm{m} \times \bm{z}) \rightarrow -\bm{\vec{\uptau}_{\parallel}} \cdot (\bm{m} \times \bm{z})$. The additional symmetry constraints caused by $M_{\textnormal{yz}}$ are
	\begin{equation} \label{torque_symmetry2}
	\begin{split}
	\uptau_{\perp}(\varphi, E)= -\uptau_{\perp}(-\varphi, E) \\
	\uptau_{\parallel}(\varphi, E)= -\uptau_{\parallel}(-\varphi, E). \\
	\end{split}
	\end{equation} 
Applying Eq. \ref{torque_symmetry1} and \ref{torque_symmetry2} to Eq. \ref{torque_form}, one obtains
	\begin{equation}
	\begin{split} \label{torque_xsymmetry}
	\uptau_{\perp} = E(A_1 \textnormal{sin}\varphi+A_3 \textnormal{sin}2\varphi+....)\\
	\uptau_{\parallel} = E(S_1 \textnormal{sin}\varphi+S_3 \textnormal{sin}2\varphi+....)
\end{split}
\end{equation}
As can be seen, $A_0$ and $S_0$ no longer exist. This result suggests that the unconventional torques would not exist in a crystal lattice having mirror planes perpendicular to both $x$ and $y$ axis. Here we only derive the case for the angle between the mirror planes are 90 degrees.   \\
To generate the unconventional SOTs, the applied in-plane current direction also matters. Eq. \ref{torque_ysymmetry} describes the case that only one mirror plane ($M_{\textnormal{xz}}$) exists and the applied current is perpendicular to $M_{\textnormal{xz}}$ plane, where unconventional SOTs are allowed. Now, let's consider the case when the lattice still only preserves the mirror plane $M_{\textnormal{xz}}$ and the current is applied parallel to $M_{\textnormal{xz}}$ (along the $x$ axis). The current does not change sign under reflection. Eq. \ref{torque_fourier} will reduce to 
\begin{equation}
\begin{split} \label{torque_xsymmetry2}
\uptau_{\perp} = E(A_2 \textnormal{cos}\varphi+A_4 \textnormal{cos}2\varphi+....)\\
\uptau_{\parallel} = E(S_2 \textnormal{cos}\varphi+S_4 \textnormal{cos}2\varphi+....)
\end{split}
\end{equation}
if the symmetry transformations are applied. Unconventional SOTs disappear in this case. Hence, when there is only one mirror plane perpendicular to the $xy$ plane, the applied in-plane current should be perpendicular to the mirror plane for generating unconventional SOTs.\\
One can also understand the impact of mirror symmetry on the SOT from the spin accumulation matrix under mirror operations. The mirror symmetry operation matrix with respect to the $xz$ plane is given by
\begin{equation*}
    M_{\textnormal{xz}}=\begin{pmatrix}
    1&0&0  \\
    0&-1&0\\
    0&0&1
    \end{pmatrix}
\end{equation*}
According to Eq. \ref{EQsym1} and  \ref{EQsym2}, $\chi$ respects:
\begin{equation}
\begin{split}
\begin{pmatrix}
    \chi_{\rm{xx}}&\chi_{\rm{xy}}&\chi_{\rm{xz}}\\
    \chi_{\rm{yx}}&\chi_{\rm{yy}}&\chi_{\rm{yz}}\\
    \chi_{\rm{zx}}&\chi_{\rm{zy}}&\chi_{\rm{zz}} 
 \end{pmatrix}  = \textnormal{det}(\textit{M}_{xz})\textit{M}_{xz}\chi \textit{M}_{xz}^T &=
  \begin{pmatrix}
    -\chi_{\rm{xx}}&\chi_{\rm{xy}}&-\chi_{\rm{xz}}\\
    \chi_{\rm{yx}}&-\chi_{\rm{yy}}&\chi_{\rm{yz}}\\
    -\chi_{\rm{zx}}&\chi_{\rm{zy}}&-\chi_{\rm{zz}} 
 \end{pmatrix}\\
& \Longrightarrow
   \chi = \begin{pmatrix}
    0&\chi_{\rm{xy}}&0\\
    \chi_{\rm{yx}}&0&\chi_{\rm{yz}}\\
    0&\chi_{\rm{zy}}&0 
 \end{pmatrix}
 \end{split}
\end{equation}

Therefore, the electric field along $x$ direction can't induce spin polarization along $z$ direction when a mirror plane with respect to the $xz$ plane presents. In contrast, the electric field along $y$ direction can induce spin polarization along $z$ direction. These conclusions drawn by using symmetry matrix analysis are consistent with ones drawn by using symmetry operation analysis. Here we only derive the case for the angle between the mirror planes are 90 degrees. Actually, no matter what the angle between these mirror planes, the unconventional SOTs would not exist if two intersecting mirror planes perpendicular to the $xy$ plane is found. Because there must be a rotation axis about the intersection line of these mirror planes. As it will be shown later that if there is a rotation axis about the $z$ axis in the lattice, the unconventional SOTs would not exist.
 
\subsection{Rotation Symmetry}
The rotation symmetries about the $x$ and $y$ axis are broken in 2D/FM heterostructures, but rotation symmetry about $z$ axis may still exist. We use the same method to analyze the impact of rotation symmetries about $z$ axis. In a lattice preserving a 2-fold rotation symmetry about the $z$ axis, the symmetry transformation leads to: (1) $\varphi \rightarrow \pi + \varphi$; (2) $E\rightarrow -E$ and (3) $\bm{\vec{\uptau}_{\perp}} \cdot \bm{z} \rightarrow \bm{\vec{\uptau}_{\perp}} \cdot \bm{z}$ and $\bm{\vec{\uptau}_{\parallel}} \cdot (\bm{m} \times \bm{z}) \rightarrow \bm{\vec{\uptau}_{\parallel}} \cdot (\bm{m} \times \bm{z})$.  Applying these symmetry transformations to Eq. \ref{torque_fourier}, one obtains,
 	\begin{equation}
	\begin{split} \label{torque_rsymmetry}
	\uptau_{\perp} = E(A_1 \textnormal{sin}\varphi+A_2 \textnormal{cos}\varphi+....)\\
	\uptau_{\parallel} = E(A_1 \textnormal{sin}\varphi+A_2 \textnormal{cos}\varphi+....)
    \end{split}
    \end{equation}      
Similar results apply to all $n$-fold ($n > $ 2) rotation symmetries. Hence, the rotation symmetry about the $z$ axis will eliminate the unconventional torque.\\
Alternatively, we can use
\begin{equation} \label{EQsym3}
\chi_{\{\textit{R}\}} = \chi+\Sigma_{(\textit{R}_i \in \{\textit{R}\})} \rm{det} (\textit{R}_i)\textit{R}_i \chi \textit{R}_i^T
\end{equation}
to directly find the $\chi$ that satisfies $\{R\}$. Besides the fact that some elements are not allowed by symmetry operations, some elements of $\chi$ are required to be same or opposite. For example, for a C$_3$ point group, 
\begin{equation}
\begin{split}
&\chi_{\rm{C_3}} = \chi+\rm{det}(\rm{C_3})\rm{C_3}\chi \rm{C_3}^T+\rm{det}[(\rm{C_3)}^2][(\rm{C_3)}^2]\chi[(\rm{C_3)}^2]^T\\
\\
&=\begin{pmatrix}
 (\frac{3}{2}(\chi_{\rm{xx}}+\chi_{\rm{yy}})&\frac{3}{2}(\chi_{\rm{xy}}-\chi_{\rm{yx}})&0\\-\frac{3}{2} (\chi_{\rm{xy}}-\chi_{\rm{yx}})&\frac{3}{2} (\chi_{\rm{xx}}+\chi_{\rm{yy}} )&0\\0&0&3\chi_{\rm{zz}}  
\end{pmatrix}  \Longrightarrow
   \chi = \begin{pmatrix}
    \chi_{\rm{xx}}&\chi_{\rm{xy}}&0\\
    \chi_{\rm{yx}}[-\chi_{\rm{xy}}]&\chi_{\rm{yy}}[=\chi_{\rm{xx}}]&0\\
    0&0&\chi_{\rm{zz}} 
 \end{pmatrix},
\end{split}
\end{equation}
where C$_3$ is the matrix for 120-degree rotation along the $z$ axis and (C$_3$)$^2$ is for 240-degree rotation. From this example, we see that we can get relations between different elements; some must be the same and some must be the opposite according to the symmetry operations.\\
We can derive $\chi$ of a specific point group by applying all symmetry operations in the set of the point group symmetry operations. In Table \ref{tab_pointgroup}, the definition of point group follows Schoenflies notation and we can obtain the current-induced spin polarization for materials with any arbitrary symmetry point group. Amorphous or polycrystalline HM/FM heterostructures can be effectively treated as C$_{\infty v}$, which is not a legitimate point group for crystalline materials according to a crystallographic restriction theorem. C$_{\infty v}$ shares the same $\chi$ with C$_{nv}$ ($n>2$) and only allows two elements: $\chi_{\textnormal{xy}}$ and $\chi_{\textnormal{yx}}$(=-$\chi_{\textnormal{xy}}$). Therefore, the current-induced spin polarization can only lie in the film plane for HM/FM heterostructures. \\

\subsection{Screw and Glide Symmetry}
\begin{figure}
    \centering
    \includegraphics[width= 16cm]{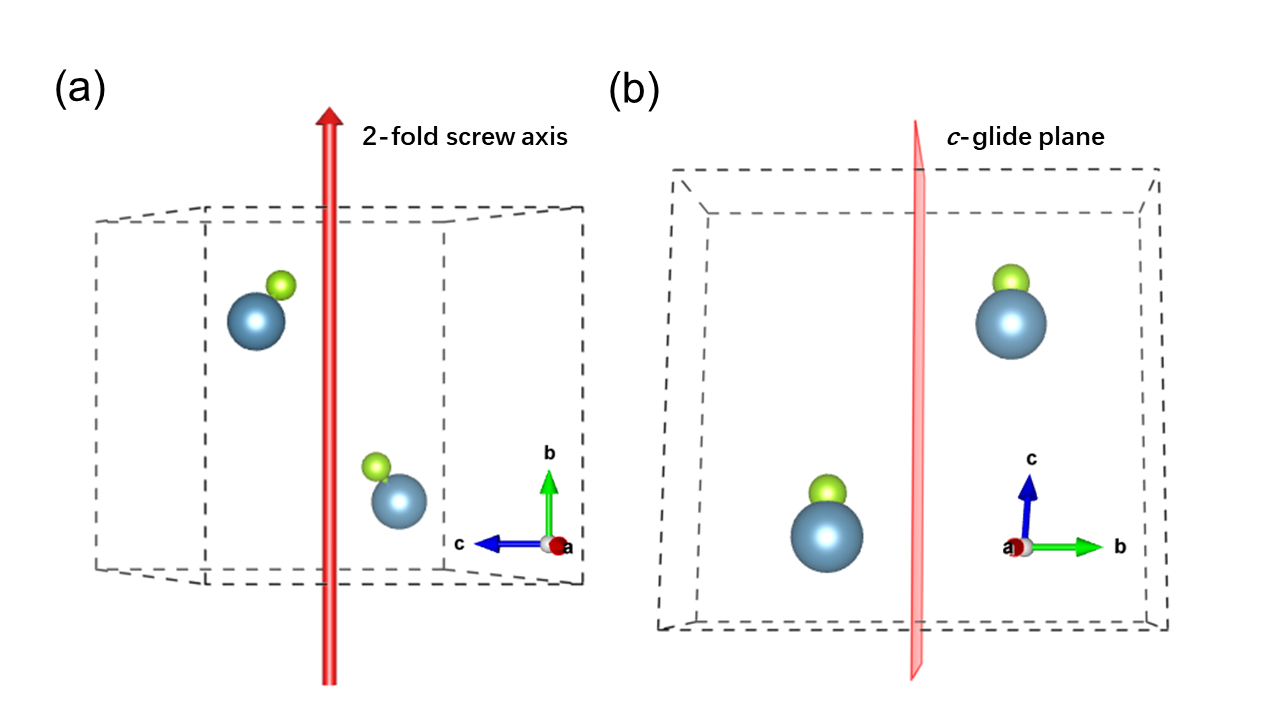}
    \caption{Schematics of screw and glide symmetry. (a) Unit cell of a lattice that preserves the 2-fold screw rotation. (b) Unit cell of a lattice that preserves the $c$-glide symmetry.}
    \label{fig:screwglide}
\end{figure}
An $n$-fold screw rotation can be decomposed of an $n$-fold rotation and a translation about the rotation axis as shown in Fig. \ref{fig:screwglide} (a). As a result, the screw symmetry about the $z$ axis in a 2D material eliminates the unconventional SOT. Fortunately, all kinds of screw symmetries are broken after interfacing to an FM, since rotation symmetries about the $x$ and $y$ axis and the translation symmetry about the $z$ axis are broken after interfacing to an FM. It is worth noting that the screw symmetry may help understand the layer dependence of current induced spin polarization\cite{macneill_thickness_2017,macneill_control_2017} which will be introduced later. \\
A glide plane can be decomposed of a mirror and a translation as shown in Fig. \ref{fig:screwglide} (b).  There are $a$-, $b$-, $c$-, $n$- and $d$- glide planes where the Latin letter represents the translation direction. Note that we first identify glide symmetries that could exist in the 2D/FM heterostructures. Whether the glide symmetry is broken or not after the 2D material interfacing to an FM can be determined by two conditions: ($i$) the glide plane is parallel to the 2D/FM interface; and ($ii$) the glide operation involves a translation along the $z$ axis. If either of the conditions is fulfilled, the glide plane symmetry is considered broken after the 2D material interfacing to an FM. Now we check every glide symmetry with these conditions. \\
Assuming the $c (a,b)$ axis of the lattice is along the $z (x,y)$ direction. Since the mirror plane perpendicular to the $z$ axis is broken after interfacing to an FM, glide planes perpendicular to $z$ axis are also broken in 2D/FM. Hence, only glide planes perpendicular to the $x$ or $y$ axis need to be considered. $c$-, $n$- and $d$- glide planes perpendicular to the $x$ or $y$ axis involve translations along the $z$ direction, hence these glide planes are also broken after interfacing to an FM. The mirror plane that is perpendicular to $x$ or $y$ axis can still be preserved after interfacing to an FM, so $a$- ($b$-) glide plane perpendicular to $y$ ($x$) axis doesn't fulfill the condition ($i$). Meanwhile, these glides planes don't involve translations along the $z$ axis, so the condition ($ii$) is also not satisfied. Therefore, $a$- ($b$-) glide plane perpendicular to $y$ ($x$) axis can be preserved after interfacing to an FM. And it can be treated as the mirror plane perpendicular to the $x$ or $y$ axis.\\  
 
\subsection{Unconventional SOT Prediction} \label{sot prediction}
\begin{figure}
    \centering
    \includegraphics[width=13.5cm]{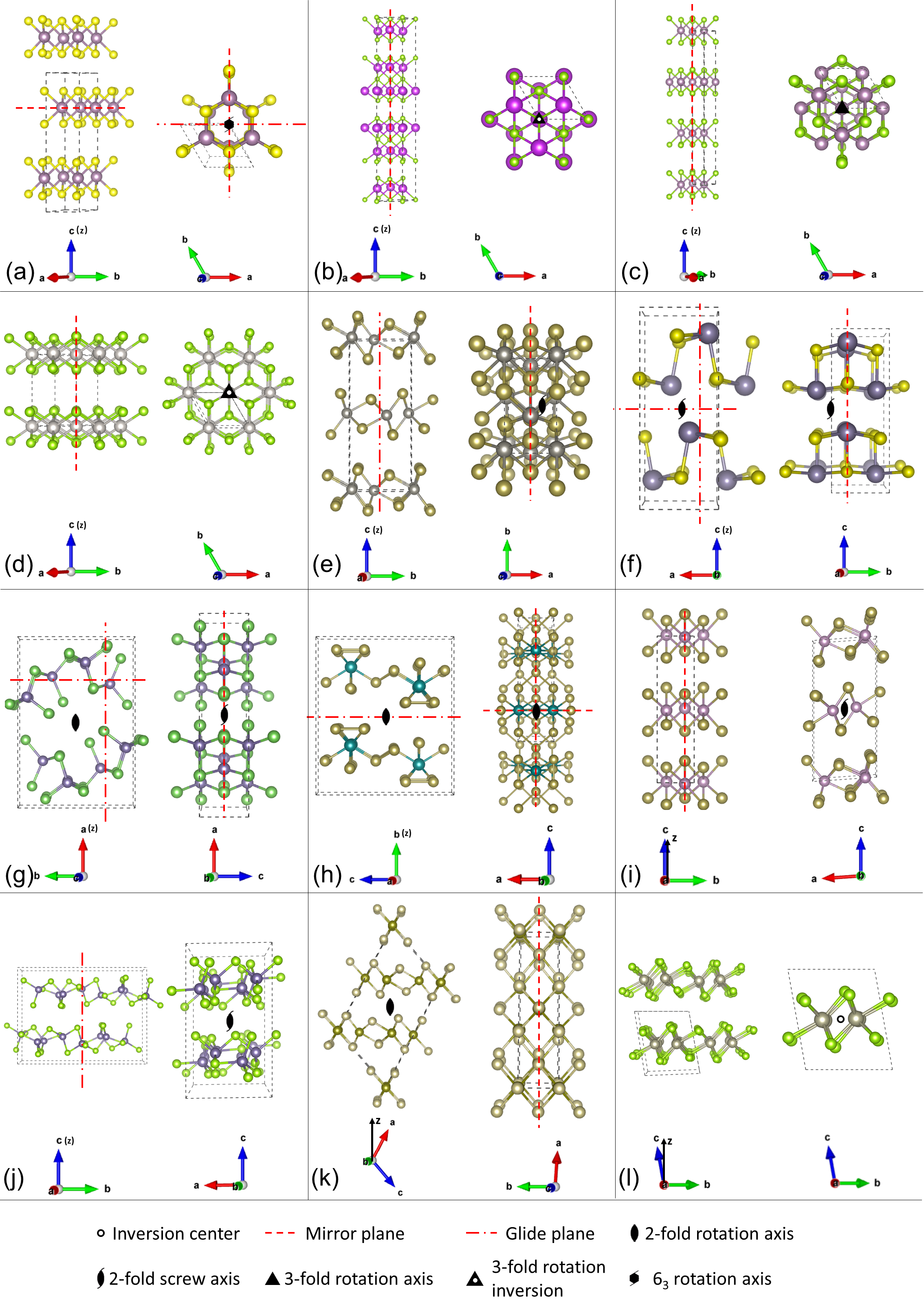}
    \caption{The crystal structure, characteristic axis and the symmetry elements lying with the characterization axis of various 2D materials. (a)MoS$_2$ (P6$_3$/mmc), (b)Bi$_2$Se$_3$ (R$\bar{3}$m), (c)MoSe$_2$ (R3m), (d)PtSe$_2$ (P$\bar{3}$m1), (e)WTe$_2$ (Pmn2$_1$), (f)SnS (Pnma), (g)GeAs$_2$ (Pbam), (h)HfTe$_5$ (Cmcm), (i)MoTe$_2$ (P2$_1$/m), (j)$\beta$ GeSe$_2$ (P2$_1$/c), (k)TaTe$_2$ (C2/m), (l)ReSe$_2$ (P$\bar{1}$) . $a$, $b$ and $c$ arrows represent the characteristic axis directions of the lattice, $z$ axis represents the surface normal direction.}
    \label{fig:my_latticestructure}
\end{figure}
 2D materials provide an opportunity to generate a desirable spin polarization thanks to their diverse symmetry properties and strong SOC. Using the above methods, we can predict if unconventional SOTs are allowed or not in 2D/FM heterostructures.
To allow unconventional SOTs the crystal lattice of 2D materials should satisfy the following conditions: ($i$) no $n$-fold rotation ($n>$1) axis about the surface normal direction ($z$ axis) exists; ($ii$) the lattice preserves at most one mirror plane or glide plane ($a$- or $b$-glide ) perpendicular to an in-plane axis. Fortunately, the needed symmetry information is already contained in the space group notation of the material. One can conveniently grab the symmetry elements about the characteristic direction of the lattice from the material's Hermann–Mauguin notation. Table \ref{table_2Dsymmetry} lists space groups for common 2D materials. The crystal structure, characteristic axis and the symmetry elements for various 2D materials are shown in Fig. \ref{fig:my_latticestructure}. In the following, we use two examples to illustrate how to determine if the unconventional SOTs are allowed. \\
First, we take 1T MoSe$_2$ with a space group of R3m and SnS with a space group of Pnma as examples. The crystal structures of SnS and MoSe$_2$ are shown in Fig. \ref{fig:my_latticestructure}(c) and (f). The full space group notation of MoSe$_2$ is R3m which has a trigonal lattice structure. As shown in Fig. \ref{fig:my_latticestructure}(c), the characteristic crystal axis $c$ is along the $z$ direction and perpendicular to the $a$ and $b$ axis. The angle between $a$ and $b$ axis is 120 degrees and they have the same symmetry operations. The space group notation shows there is a 3-fold rotation axis about the $c$ axis so that condition ($i$) is not satisfied. Besides, there are two mirror planes perpendicular to a and b axis, respectively. Therefore, condition ($ii$) is not satisfied either. For its $a$ and $b$ axis, there is a mirror plane perpendicular to them, respectively. Hence, unconventional SOTs are not allowed in 1T MoSe$_2$ with a space group of R3m.\\
\begin{table}
    \centering
    	\caption{The full space group notation of various 2D materials, the point group for the 2D material (2D/FM) and unconventional SOT predictions.}
			\label{table_2Dsymmetry}
			\resizebox{\textwidth}{!}{
		\begin{tabular}{p{5cm}ccc}
			\hline \hline \\
			2D materials&Space group of 2D&2D/FM&Unconventional $\uptau_{\textnormal{DL}}$\\ \hline
			Black Phosphrous&C2/m 2/c 2$_1$/a\cite{li_emerging_2019}& C$_{2v}$ &N\\
			SnS, SnSe, GeS, GeSe&P2$_1$/n 2$_1$/m 2$_1$/a\cite{li_emerging_2019,zhao_-plane_2020}& C$_{1v}$&Y\\
			GeAs$_2$&P2$_1$/b 2$_1$/a 2/m\cite{li_emerging_2019}& C$_{1v}$ &Y\\
			SiP&C2/m 2/c 2$_1$\cite{zhao_-plane_2020}& C$_{1v}$ &Y\\
			GaP, GeAs, GeTe&C1 2/m 1\cite{li_emerging_2019,zhao_-plane_2020}& C$_{1v}$ &Y\\
			GeS$_2$, $\beta$-GeSe$_2$, MoO$_2$&P1 2$_1$/c 1\cite{li_emerging_2019}& C$_1$ &Y\\
			TiS$_3$, TiSe$_3$,ZrS$_3$, ZrSe$_3$, HfS$_3$, HfSe$_3$&P1 2$_1$/m 1\cite{li_emerging_2019,zhao_-plane_2020}&C$_1$ &Y\\
			2H-NbSe$_2$, MoS$_2$, MoSe$_2$, WS$_2$, WSe$_2$, MoTe$_2$, SnSe$_2$&P6$_3$/m 2/m 2/c\cite{saito_raman_2016}& C$_{3v}$ &N \\
			1T-MoS$_2$, MoSe$_2$, WS$_2$, WSe$_2$, MoTe$_2$, SnSe$_2$ &P$\bar{3}$ 2/m 1\cite{saito_raman_2016}& C$_{3v}$ &N \\
			1T’-MoS$_2$, MoSe$_2$, WS$_2$, WSe$_2$, MoTe$_2$, SnSe$_2$&P1 2$_1$/m 1\cite{saito_raman_2016}&C$_{1v}$ &Y \\
			3R-MoS$_2$, MoSe$_2$, WS$_2$, WSe$_2$, MoTe$_2$, SnSe$_2$&R3m\cite{saito_raman_2016}& C$_{3v}$ &N \\
            1T’-ReSe$_2$, ReS$_2$&P$\bar{1}$\cite{saito_raman_2016}& C$_1$ &Y \\ 
            T$_\textnormal{d}$-MoS$_2$, T$_\textnormal{d}$-MoTe$_2$, T$_\textnormal{d}$-WTe$_2$, TaIrTe$_4$&Pmn2$_1$\cite{macneill_control_2017,zhao_-plane_2020}& C$_{1v}$ &Y\\
            Bi$_2$Se$_3$, Bi$_2$Te$_3$, Sb$_2$Te$_3$, Sb$_2$Se$_3$&R$\bar{3}$ 2/m\cite{zhang_topological_2009}&C$_{3v}$ &N\\
            1T’-TaTe$_2$, NbTe$_2$&C1 2/m 1\cite{stiehl_current-induced_2019,brown_crystal_1966}&C$_{1v}$ &Y\\
            PtSe$_2$, PtTe$_2$, PdTe$_2$, PtBi$_2$&P$\bar{3}$ 2/m 1\cite{huang_type-ii_2016}& C$_{3v}$ &N\\
            ZrTe$_5$, HfTe$_5$, Ta$_2$NiS$_5$&C2/m 2/c 2$_1$/m\cite{weng_transition-metal_2014}&C$_{2v}$ &N\\
            TlSe&I4/m 2/c 2/m\cite{zhao_-plane_2020}&C$_{2v}$&N\\
			\hline \hline \\
		\end{tabular}
	}
\end{table}
The full space group notation of SnS is P2$_1$/n2$_1$/m2$_1$/a,  the space group notation implies that the crystal has an orthorhombic lattice structure. The characteristic crystal directions $a$, $b$ and $c$ are perpendicular to each other as shown in Fig. \ref{fig:my_latticestructure}(f). About the $c$ axis ($z$ axis), 2$_1$/a suggests that there are a 2-fold screw axis and an $a$-glide plane perpendicular to it. About the $a$ axis ($x$ axis), 2$_1$/n suggests that there are a 2-fold screw axis and an $n$-glide plane perpendicular to it. About the $b$ axis ($y$ axis), 2$_1$/m suggests that there are a 2-fold screw axis and a mirror plane perpendicular to it. We need to check conditions ($i$) and ($ii$) to determine if unconventional SOTs are allowed. For SnS, the crystal $c$ axis is along $z$ axis and there is no rotation axis along $c$ axis as indicated by the space group notation. Condition ($i$) is satisfied. In SnS, only one mirror plane perpendicular to the $b$ axis as indicated by the space group notation. Condition ($ii$) is satisfied. Hence, unconventional SOTs are allowed in SnS by symmetry. \\
The symmetry matrix analysis is very useful for theoretical calculations. To predict unconventional SOTs with the symmetry matrix, the point group of the 2D/FM heterostructure needs to be determined since the integration could break some symmetry operations of the 2D materials. We then obtain the corresponding  magnetoelectric pseudovector using Table \ref{tab_pointgroup} according to the point group. Table \ref{table_2Dsymmetry} summarizes the point groups for various 2D/FM bilayers and our predictions about the existence of unconventional SOTs. \\  

\section{Review of the Experimental Progress}
Since the inversion symmetry in 2D/FM heterostructures is broken, we expect an in-plane spin accumulation, an FL-SOT in the form of $\bm{{\uptau_{\textnormal{FL}}}} \propto \bm{m} \times (\bm{z} \times \bm{j})$ and a DL-SOT in the form of $\bm{\uptau_{\textnormal{DL}}} \propto \bm{m} \times [\bm{m} \times (\bm{z} \times \bm{j})]$. As argued above, an out-of-plane spin polarization can be expected in heterostructures with lateral symmetry breaking. As a result, unconventional DL- and FL-SOTs could be found. To compare experiment results from the literature, we use SOT efficiency ($\xi$) and spin conductivity ($\sigma$), which are defined by SOT efficiency $\xi = (2e/\hbar) J_{\textnormal{S}}/J_{\textnormal{C}}$ ($J_{\textnormal{S}}$: spin current density, $J_{\textnormal{C}}$: charge current density) and spin conductivity $\sigma = J_{\textnormal{S}}/E$ ($E$: electric field), respectively. These two values differ by a factor of the electrical conductivity ($\sigma_{\textnormal{C}}$) of the SOC material: $\xi = (2e/\hbar) \sigma /\sigma_{\textnormal{C}}$. SOT efficiencies of conventional FL-SOT ($\bm{{\uptau_{\textnormal{A}}}}$), DL-SOT ($\bm{{\uptau_{\textnormal{S}}}}$), unconventional FL-SOT ($\bm{{\uptau_{\textnormal{T}}}}$), and DL-SOT ($\bm{{\uptau_{\textnormal{B}}}}$) are expressed as $\xi_{\textnormal{A}}$, $\xi_{\textnormal{S}}$, $\xi_{\textnormal{T}}$, and $\xi_{\textnormal{B}}$, respectively. Their spin conductivities are expressed as $\sigma_{\textnormal{A}}$, $\sigma_{\textnormal{S}}$, $\sigma_{\textnormal{T}}$, and $\sigma_{\textnormal{B}}$, respectively. \\


\subsection{Topological Insulators}
In search of materials to provide efficient SOTs,  TIs such as the Bi$_2$Se$_3$ family are among the most promising candidates owing to its helical spin structure (spin-momentum locking) of the topological surface states (TSSs).  Earlier reviews that mention or focus on spintronics based on TIs include refs..\cite{pesin_spintronics_2012, fan_spintronics_2016,wang_fmr-related_2018,manchon_current-induced_2019} Here, we summarize recent experimental progress of TI-based SOT devices that are highly tunable or compatible with CMOS technology. It is worth mentioning that the discussion in this section is limited to conventional SOTs. Because unconventional SOTs in these materials are not allowed by symmetry according to the previous symmetry analysis as shown in Table \ref{table_2Dsymmetry}.  Note that this review focuses on TIs that are 2D materials. \\
\begin{figure}
	\centering
	\includegraphics[width=16cm]{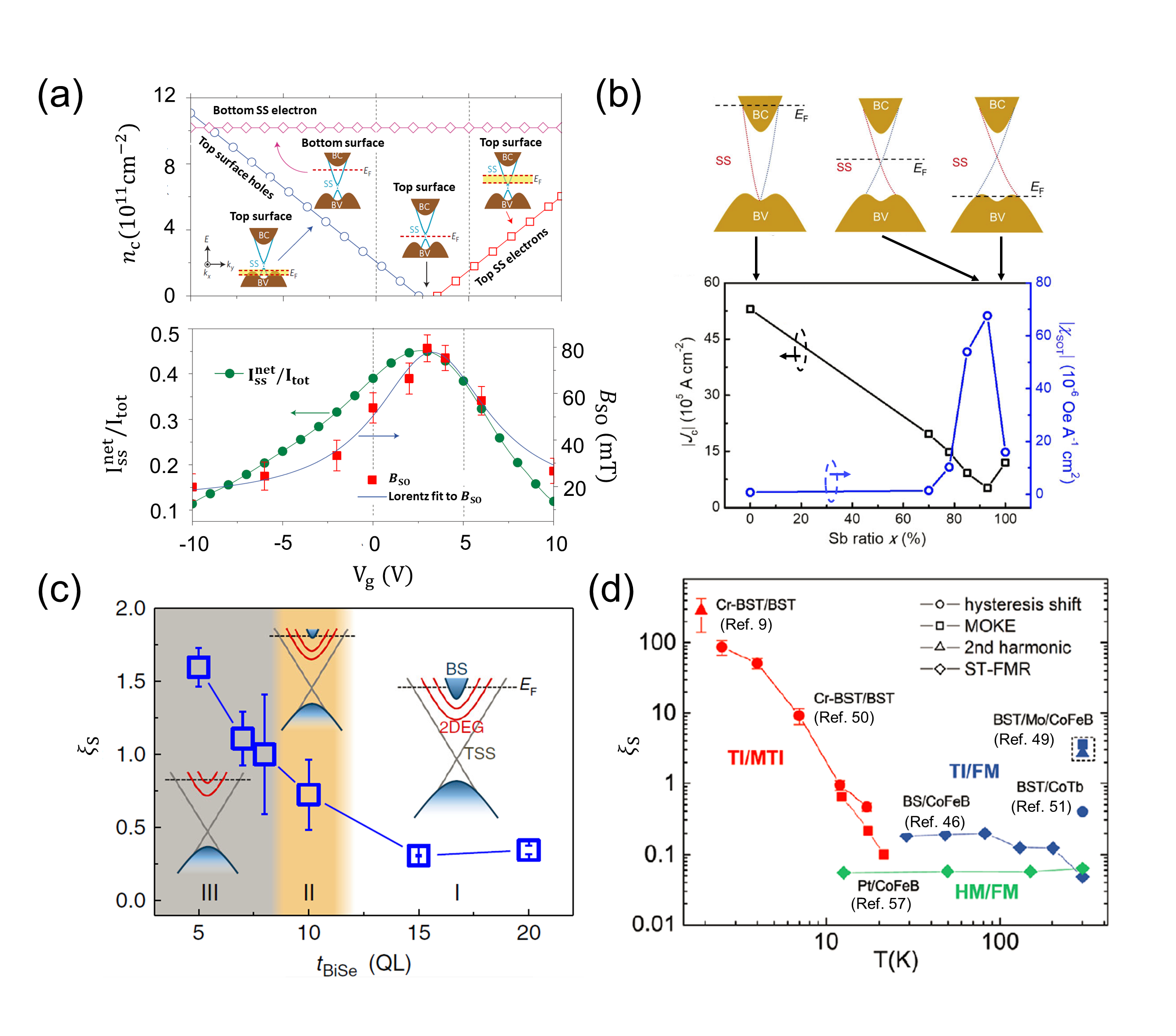}
	\vspace{0pt}
	\caption{Tuning SOTs from TIs through electric field, composition, thickness, and temperature. (a) Control the Fermi level and the carrier density $n_\textnormal{C}$ by electric fields. The upper panel shows the top and bottom surface carrier densities as functions of gate voltage V$_{\textnormal{g}}$. The lower panel shows the ratio between the net spin-polarized surface current over total current and the effective spin orbit field $B_{\textnormal{SO}}$ as functions of V$_{\textnormal{g}}$.  (b) The SOT effective field $\chi_{\textnormal{SOT}}$ and switching current density as a function of Sb content. The Fermi level is indicated by the upper panel insets. (c) The SOT efficiency ($\xi_{\textnormal{S}}$) as a function of Bi$_2$Se$_3$ thickness at room temperature. The inset shows the schematic of the band structure.(d) The $\xi_{\textnormal{S}}$ as a function of temperature for various devices. The figures are reproduced from Fan $et$ $al$., \cite{fan_electric-field_2016} Wu $et$ $al$.,\cite{wu_room-temperature_2019} Wang $et$ $al$.\cite{wang_room_2017} and Che $et$ $al$..\cite{che_strongly_2020}}
	\label{fig:efti}
\end{figure}
\textbf{Tuning SOTs from TIs.}  Carrier density of TSSs plays a crucial role in determining the SOT efficiency in a TI-based heterostructure. Experimentally, Fan $et$ $al$.\cite{fan_electric-field_2016} have shown that the SOT strength can be tuned by a factor of four in a Cr-doped TI device with a top gate voltage. The TI device shows the largest SOT efficiency when the carrier density of the top surface is close to the minimum and the overall surface state carrier density reaches the maximum as shown in Fig. \ref{fig:efti} (a). Tuning carrier density of TSSs by varying the Bi:Sb ratio in (Bi$_{1-x}$Sb$_x$)$_2$Te$_3$ has been shown by Wu $et$ $al$..\cite{wu_room-temperature_2019} As shown in Fig. \ref{fig:efti} (b), the maximum SOT effective field is also found when the Fermi level is close to the Dirac point and the transport signatures are dominated by the TSSs. Note that Kondou $et$ $al$.\cite{kondou_fermi-level-dependent_2016} showed that the SOT efficiency reaches the minimum when the Fermi level is close to the Dirac point, which was attributed to possible inhomogeneity of Dirac electron momentum and/or instability of the helical spin structure.  \\
SOTs from TIs depend on the layer thickness. Wang $et$ $al$.\cite{wang_room_2017} have studied the SOT efficiency as a function of the TI thickness in a Bi$_2$Se$_3$/Py heterostructure as shown in Fig. \ref{fig:efti} (c). The SOT efficiency increases with the decrease of TI thickness, resulting from decreasing bulk states and 2DEG states and increasing TSSs.  \\
The SOT efficiency of TI based device is sensitive to the temperature. Che $et$ $al$.\cite{che_strongly_2020} studied the temperature dependence of SOT of a TI/MTI device using the hysteresis shift and MOKE methods. The TI shows a much larger SOT efficiency when the temperature is below 12 K which is attributed to a higher spin polarization ratio of the TSS at low temperatures. The temperature dependence of spin Hall angle for various TI based devices is shown in Fig. \ref{fig:efti} (d). It can be seen that the SOT efficiency of HM/FM device has a weak dependence on the temperature,\cite{wang_determination_2014} whereas the SOT efficiency of TI/MTI and TI/FM devices shows much higher values at low temperatures. \\
\begin{figure}
    \centering
    \includegraphics[width=16cm]{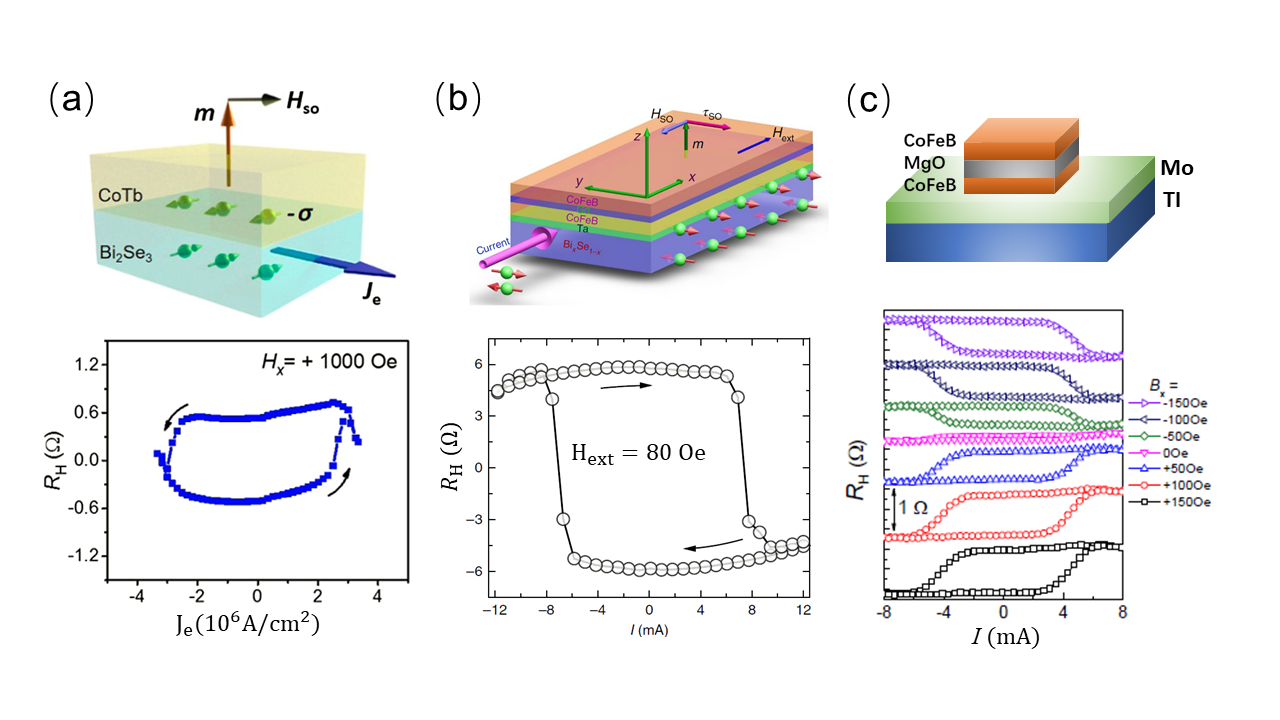}
    \caption{Current induced magnetization switching in TI/FM stacks with PMA: (a) Bi$_2$Se$_3$/CoTb, (b) Bi$_x$Se$_{1 - x}$/Ta(0.5nm)/CoFeB/Gd/CoFeB, and (c) (BiSb)$_2$Te$_3$/Mo(2nm)/CoFeB stacks. The figures are reproduced from Han $et$ $al$. \cite{han_room-temperature_2017}, Mahendra $et$ $al$. \cite{dc_room-temperature_2018} and Shao $et$ $al$..\cite{shao_room_2018}}
    \label{fig:TIPMA}
\end{figure}
\textbf{Room temperature perpendicular magnetization switching using TIs.} A major challenge is to integrate a ferromagnetic layer with PMA on TI. Han $et$ $al$. \cite{han_room-temperature_2017} have shown SOT switching of perpendicular magnetized materials in Bi$_2$Se$_3$/ and (BiSb)$_2$Te$_3$/CoTb utilizing the strong bulk PMA of CoTb at room temperature as shown in Fig. \ref{fig:TIPMA} (a).\\
The PMA can also be achieved by inserting a metal layer between TI and a ferromagnetic layer. It is known that the Ta seed layer can promote PMA in CoFeB thin films since it absorbs boron increasing the crystallization of CoFeB. Mahendra $et$ $al$. \cite{dc_room-temperature_2018} have obtained PMA in TI based heterostructures by inserting a thin Ta layer between Bi$_x$Se$_{1-x}$ and CoFeB stack as shown in Fig. \ref{fig:TIPMA} (b). The required current density to switch the magnetization is only around 4.3 $\times$ 10$^5$ Acm$^{-2}$. However, the drawback of inserting a heavy metal layer between TI and an FM layer is that the insertion layer hinders the spin current diffusion to the FM layer. To maintain a high SOT efficiency, inserting a light metal with a weak SOC strength could be a better option. Shao $et$ $al$. \cite{shao_room_2018} have shown current induced magnetization switching in Bi$_2$Se$_3$/, Bi$_2$Te$_3$/ and (BiSb)$_2$Te$_3$/Mo/CoFeB structures with PMA. The switching current for the (BiSb)$_2$Te$_3$/Mo/CoFeB device is as low as 3$\times$10$^5$Acm$^{-2}$ and a large SOT efficiency of about 2.66 is obtained.  The advantages of inserting a Mo layer are threefold: first, Mo has a weak SOC and does not signficantly reduce the SOT efficiency; second, Mo/CoFeB/MgO has a strong PMA at room temperature; third, Mo has a high thermal stability compared with Ta and thus is more compatible with modern CMOS technologies that require annealing temperature as high as $\ang{400}$C. \\
To utilize TIs for future SOT-based devices, two important directions need to be pursued. First, since unconventional SOTs are not allowed in TI/FM heterostructures, methods to achieve reliable field-free switching for TI-based devices need to be investigated. Potential methods include a lateral structure asymmetry,\cite{yu_switching_2014} tilted anisotropy,\cite{you_switching_2015} and exchange bias.\cite{fukami_magnetization_2016,oh_field-free_2016} Second, since the high resistance of TIs could compensate for the advantage of TIs’ high SOT efficiency,\cite{li_materials_2020} conductive TIs with giant SOTs are highly desirable. Promising results have been shown recently that a conductive TI Bi$_{0.9}$Sb$_{0.1}$ was reported to host a very large SOT efficiency with a very low resistivity.\cite{khang_conductive_2018} \\
\begin{table}
	\caption{\label{tab_SOTTI} The SOT efficiencies and spin conductivities of TI/FM heterostructures.}
	\resizebox{\textwidth}{!}
	{
		\begin{tabular}{lccccl}
			\hline \hline
			\multirow{2}{*}{Authors}&\multirow{2}{*}{Materials}&Characterization&\multirow{2}{*}{$\xi$}&{$\sigma$}&\multirow{2}{*}{Note}\\
			&&technique&&($\times$ 10$^3$ ($\hbar/2e$)($\Omega m)^{-1}$)&\\ \hline
			\multirow{2}{*}{Fan $et$ $al$. \cite{fan_magnetization_2014}}&\multirow{2}{*}{(BiSb)$_2$Te$_3$/Cr-(BiSb)$_2$Te$_3$} &\multirow{2}{*}{SHH}&$\xi_{\textnormal{S}}$=140-425&$\sigma_{\textnormal{S}}$=1540-4675&measured at 1.9 K and \\
			&&&$\xi_{\textnormal{A}}$=26&$\sigma_{\textnormal{A}}$=286&magnetization angle dependent\\\hline
			\multirow{2}{*}{Che $et$ $al$. \cite{che_strongly_2020}}&\multirow{2}{*}{(BiSb)$_2$Te$_3$/Cr-(BiSb)$_2$Te$_3$}
			&Hysteresis shift&\multirow{2}{*}{$\xi_{\textnormal{S}}$=0.1-100}&&\multirow{2}{*}
			{temperature dependent}\\
			&&MOKE&&&\\\hline
			\multirow{2}{*}{Mellnik $et$ $al$. \cite{mellnik_spin-transfer_2014}}&\multirow{2}{*}{Bi$_2$Se$_3$/Py}&\multirow{2}{*}{ST-FMR}
			&$\xi_{\textnormal{S}}$=2.0-3.5&$\sigma_{\textnormal{S}}$=110-200&thickness and device \\
			&&&$\xi_{\textnormal{A}}\approx$2.5&$\sigma_{\textnormal{A}}\approx$150&dependent\\\hline
			\multirow{2}{*}{Han $et$ $al$. \cite{han_room-temperature_2017}}&Bi$_2$Se$_3$/CoTb&\multirow{2}{*}{Hysteresis shift}&
			$\xi_{\textnormal{S}}$=0.16&$\sigma_{\textnormal{S}}$=15&\\
			&(Bi,Sb)$_2$Te$_3$/CoTb&&$\xi_{\textnormal{S}}$=0.40&$\sigma_{\textnormal{S}}$=10&\\\hline
			Wang $et$ $al$. \cite{wang_room_2017}&Bi$_2$Se$_3$/CoFeB&ST-FMR&0.3-1.6&& thickness dependent\\\hline
			\multirow{2}{*}{Wang $et$ $al$. \cite{wang_topological_2015}}&\multirow{2}{*}{Bi$_2$Se$_3$/CoFeB}&\multirow{2}{*}{ST-FMR}
			&$\xi_{\textnormal{S}}$=0.047-0.42&$\sigma_{\textnormal{S}}$=6.3-93&\multirow{2}{*}
			{temperature dependent}\\
			&&&$\xi_{\textnormal{A}}$=0-0.33&$\sigma_{\textnormal{A}}$=0-73&\\\hline
			\multirow{4}{*}{Mahendra $et$ $al$. \cite{dc_room-temperature_2018}} &\multirow{2}{*}{Bi$_x$Se$_{1-x}$/CoFeB}&SHH&$\xi_{\textnormal{S}}$=0.45-18.62
			&$\sigma_{\textnormal{S}}$=145&\\
			&&ST-FMR&$\xi_{\textnormal{S}}$=1.56-8.67&&thickness dependent\\
			&\multirow{2}{*}{Bi$_x$Se$_{1-x}$/Ta/CoFeB}&SHH&$\xi_{\textnormal{S}}$=6&&and poly-crystalline\\
			&&ST-FMR&$\xi_{\textnormal{S}}$=1.35&&\\\hline
			\multirow{5}{*}{Shao $et$ $al$. \cite{shao_room_2018}}&Bi$_2$Se$_3$/CoFeB&SHH&$\xi_{\textnormal{S}}$=0.35
			&$\sigma_{\textnormal{S}}$=32&\\
			&Bi$_2$Te$_3$/CoFeB&SHH&$\xi_{\textnormal{S}}$=1.76&$\sigma_{\textnormal{S}}$=147&\\
			&(BiSb)$_2$Te$_3$/CoFeB&SHH&$\xi_{\textnormal{S}}$=8.33&$\sigma_{\textnormal{S}}$=146&\\
			&\multirow{2}{*}{(BiSb)$_2$Te$_3$/Mo/CoFeB}&SHH&
			\multirow{2}{*}{$\xi_{\textnormal{S}}$=2.66}&	\multirow{2}{*}{$\sigma_{\textnormal{S}}$=106}&\\
			&&MOKE&&&\\\hline
			Khang $et$ $al$.\cite{khang_conductive_2018}&Bi$_{0.9}$Sb$_{0.1}$/Mn$_{0.4}$Ga$_{0.6}$
			&Coercivity change&
			$\xi_{\textnormal{S}}$=52&$\sigma_{\textnormal{S}}$=13000&non-van der Waals material\\
			\hline\hline
		\end{tabular}
	}
\end{table}\\

\subsection{Transition Metal Dichalcogenides}
Earlier reviews about spintronics based on 2D materials focus on spin injection, transport and detection, \cite{han_graphene_2014,han_perspectives_2016} and spin-valley coupling.\cite{schaibley_valleytronics_2016} Recently, Lin $et$ $al$.\cite{lin_two-dimensional_2019} reviewed functional spintronic devices and circuits based on 2D materials. Current-induced spin polarization in 2D materials and the associated SOT effects have not been reviewed. Here, we focus on SOTs from TMDs and show how the symmetry arguments align with the experimental observations.\\
\textbf{Experimental observations of SOTs from TMDs.} 2H-MoS$_2$ is an intensively studied 2D material. It has a hexagonal lattice structure and a space group of P6$_3$/mmc with each Mo atom bonded with 6 S atoms, as illustrated by Fig. \ref{fig:my_latticestructure} (a). Monolayer MoS$_2$ has a space group of P6/mmc since there is no 3-fold screw symmetry along the c-axis.  When interfacing to an FM layer, the structure of the materials fulfills the symmetry requirements to generate the conventional $\uptau_{\textnormal{DL}}$ and  $\uptau_{\textnormal{FL}}$ due to the symmetry breaking at the interface. However, neither monolayer nor multilayer MoS$_2$/FM heterostructures fulfill the symmetry requirements to generate unconventional SOTs, since there is more than one mirror plane perpendicular to an in-plane axis as illustrated by Fig. \ref{fig:my_latticestructure} (a).\\
\begin{figure}
    \centering
    \includegraphics[width=16cm]{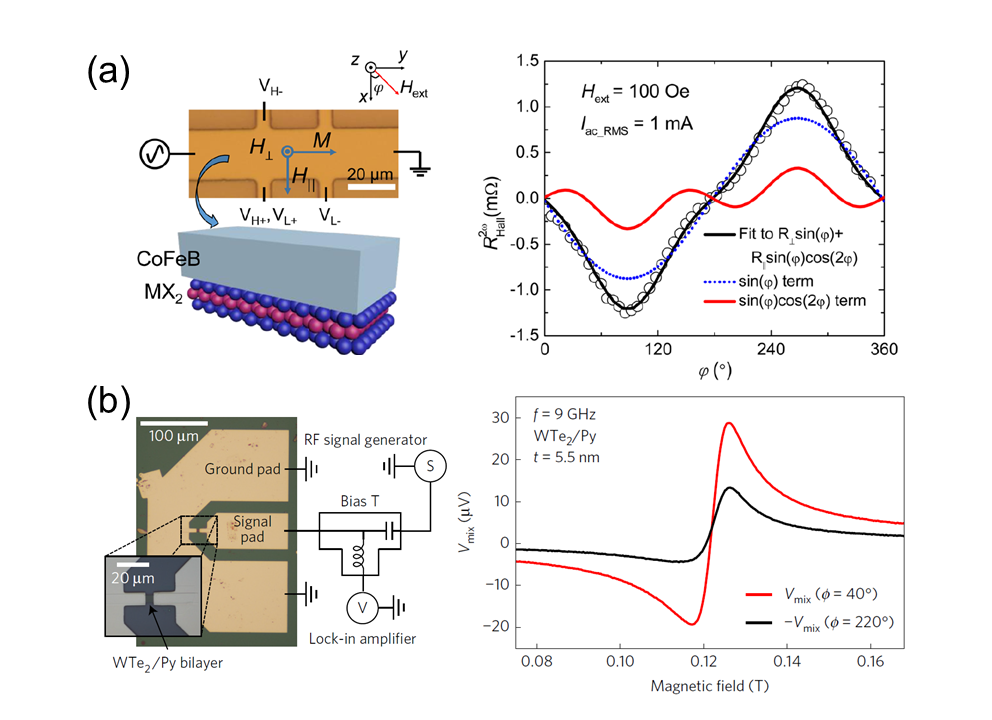}
    \caption{The measurement setup and results of SOTs from TMDs. (a) Measurement setup of SOT measurements for the MX$_2$/CoFeB heterostructures and the second-harmonic Hall resistance as a function of in-plane azimuthal angle ($\varphi$) with an external magnetic field 100 Oe applied. (b) The ST-FMR measurement setup on the SOT of WTe$_2$/Py heterostructures and ST-FMR signals for a WTe$_2$(5.5 nm)/Py(6 nm) sample with current applied along the a-axis. $\phi$ is angle between the current and magnetization. The figures are adapted from Shao $et$ $al$. \cite{shao_strong_2016} and Macnei $et$ $al$.. \cite{macneill_control_2017}}
    \label{fig:mos2}
\end{figure}\\
Shao $et$ $al$. \cite{shao_strong_2016} reported the SOTs generated by monolayer MoS$_2$ and WSe$_2$ with CoFeB at room temperature, as illustrated in Fig.\ref{fig:mos2} (a), using the SHH method. Different SOTs can be distinguished by analyzing different contributions to the azimuthal angle ($\varphi$) dependence of the second harmonic Hall resistance ($R_{\textnormal{H}}^{2\omega}$). As shown in Fig.\ref{fig:mos2} (a), the current is applied along the y-axis and $\varphi$ is the angle between magnetization and the x-axis. If the magnetization of the FM layer is always in the plane, for an in-plane current, the potentially generated conventional $\uptau_{\textnormal{A}}$ and unconventional $\uptau_{\textnormal{B}}$ will be out-of-plane and the conventional $\uptau_{\textnormal{S}}$ and unconventional $\uptau_{\textnormal{T}}$ will be in-plane. The in-plane SOTs modulate the anomalous Hall resistance that is independent of $\varphi$. The out-of-plane SOTs modulate the planar Hall resistance, contributing to a cos(2$\varphi$) dependence. $\uptau_{\textnormal{A}}$ and $\uptau_{\textnormal{S}}$ originate from the in-plane spin polarization, contributing to a sin$\varphi$ dependence. $\uptau_{\textnormal{B}}$ and $\uptau_{\textnormal{T}}$ originate from the out-of-plane spin polarization that is independent of $\varphi$. In summary, SOT contributions to $R_{\textnormal{H}}^{2\omega}$ can be fitted to the first order using 
\begin{equation*}
    R_{\textnormal{H}}^{2\omega} (\varphi) = A\textnormal{sin}(\varphi)\textnormal{cos}(2\varphi)+B\textnormal{cos}(2\varphi)
    +S\textnormal{sin}(\varphi)+T,
\end{equation*}
where $A$, $B$, $S$, and $T$ correspond to the contributions from $\uptau_{\textnormal{A}}$, $\uptau_{\textnormal{B}}$, $\uptau_{\textnormal{S}}$, and $\uptau_{\textnormal{T}}$, respectively. The typical azimuthal angle dependence of $R_{\textnormal{H}}^{2\omega} (\varphi)$ of MoS$_2$/CoFeB is shown in Fig.\ref{fig:mos2} (a).\cite{shao_strong_2016} Only the conventional FL-SOT was observed in this experiment, where $\sigma_{\textnormal{A}}$ is 2.9 $\times$ 10$^3$ ($\hbar/2e$)($\Omega \textnormal{m})^{-1}$ for MoS$_2$ and 5.5 $\times$ 10$^3$ ($\hbar/2e$)($\Omega \textnormal{m})^{-1}$ for WSe$_2$. Although a sin$\varphi$ dependence was observed (Fig.\ref{fig:mos2} (a)), it is due to thermoelectric effect instead of SOT effect.\cite{shao_strong_2016} The observed FL-SOT is attributed to the Rashba-Edelstein effect since it is much larger than the DL-SOT. The generated SOTs can be different if the FM layer is different. Zhang $et$ $al$.\cite{zhang_research_2016} reported the SOTs in a monolayer MoS$_2$/Py device. The observed symmetric ST-FMR peak is about 4 times bigger than the antisymmetric peak in a MoS$_2$/Py heterostructure, which suggests that the DL-SOT could be much larger than the FL-SOT according to conventional ST-FMR analysis.\cite{liu_spin-torque_2011} However, the strength of SOT was not quantified and the origin of SOT was not interpreted due to a potentially large contribution from inverse Rashba-Edelstein effect-induced spin pumping.\\ 
\begin{figure}
	\centering
	\includegraphics[width=16cm]{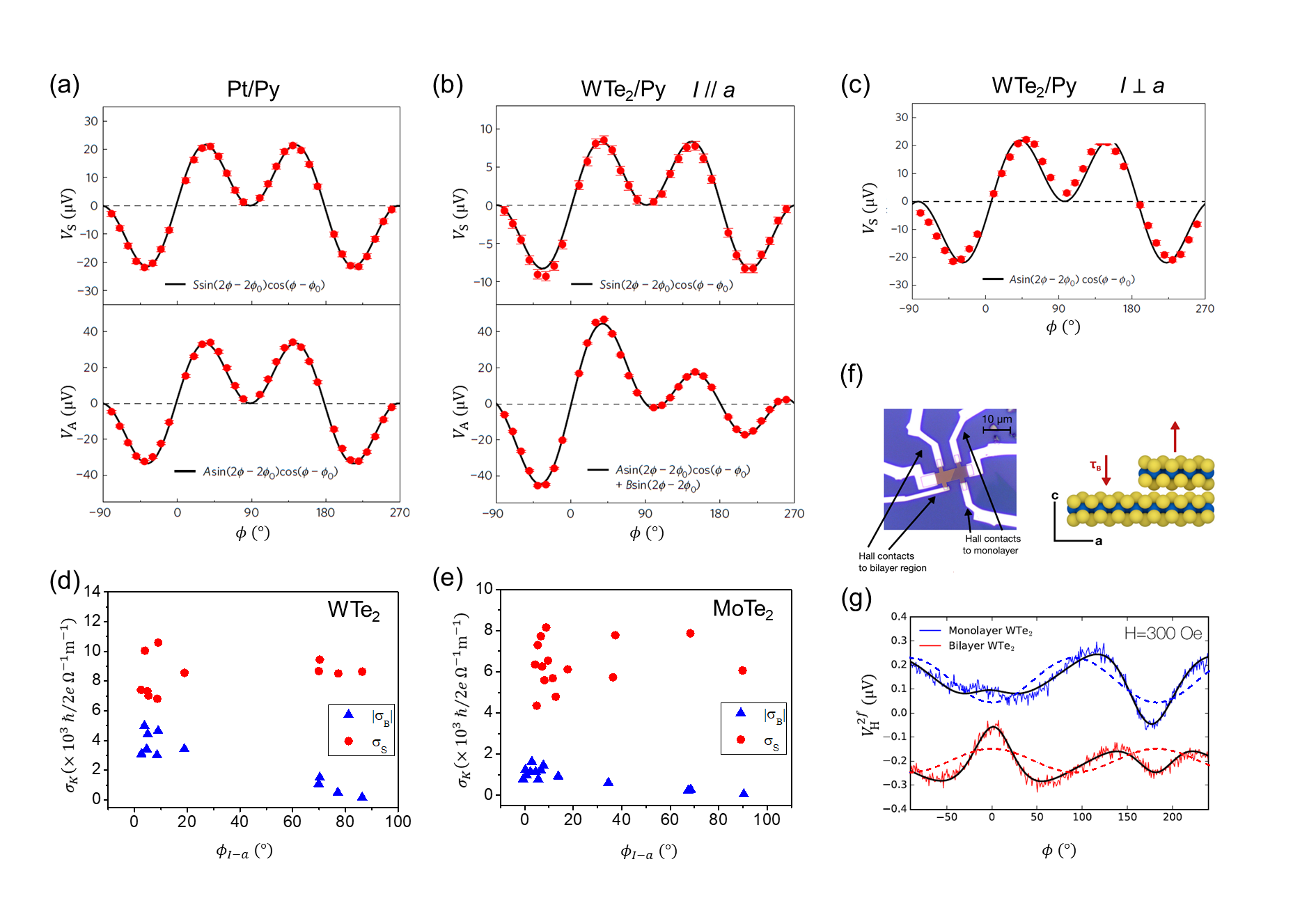}
	\vspace{0pt}
	\caption{The angular dependence of ST-FMR signals of TMD/Py samples. Azimuthal angle dependence of ST-FMR resonance components for (a) a Pt(6 nm)/Py (6 nm) device, and for a WTe$_2$(5.5 nm)/Py (6 nm) device with (b) current applied parallel to the $a$ axis and (c) current applied perpendicular to the $a$ axis. $\phi$ is the angle between the applied current and magnetization. (d) and (e)  $\uptau_{\textnormal{B}}$ and $\uptau_{\textnormal{S}}$  as a function of the angle ($\phi_{I-a}$) between the applied current and the $a$-axis for WTe$_2$ and MoTe$_2$, respectively. (f)The optical micrograph of a WTe$_2$/Py Hall bar device and schematic of the crystal structure of WTe$_2$, showing that the surface structure is rotated by 180 degrees across a monolayer step. (g) The second harmonic Hall resistance of the monolayer and bilayer regions. The dash line shows the contribution from the unconventional damping-like SOT. $\phi$ is the angle between the applied current and magnetization. (a), (b) and (c) are reproduced from MacNeil $et$ $al$.,\cite{macneill_control_2017} (d), (f) and (g) are reproduced from MacNeil $et$ $al$..\cite{macneill_thickness_2017} (e) is reproduced from Stiehl $et$ $al$.. \cite{stiehl_layer-dependent_2019}  }
	\label{fig:wte2}
\end{figure}\\
T$_\textnormal{d}$-WTe$_2$ has a lower crystal symmetry than 2H-MoS$_2$. It has a primitive lattice which belongs to the C$_{2v}$ point group (space group Pmn2$_1$). There are a mirror plane perpendicular to its $a$ axis, an $n$-glide plane perpendicular to its $b$ axis, and a 2-fold screw axis along its $c$ axis as illustrated in Fig. \ref{fig:my_latticestructure} (e). When combined with a ferromagnetic layer such as Py, the 2-fold screw symmetry along $c$ axis and the $n$-glide symmetry along $b$ axis are broken. This lateral symmetry breaking may lead to the generation of the out-of-plane $\uptau_{\textnormal{DL}}$, which corresponds to unconventional $\bm{{\uptau_{\textnormal{B}}}}$. MacNeill $et$ $al$. \cite{macneill_control_2017} reported SOTs in a WTe$_2$/Py heterostructure as shown in Fig. \ref{fig:mos2} (b). The SOTs were measured by the ST-FMR method. The field dependence of the ST-FMR signal is decomposed into anti-symmetric and symmetric Lorentzian parts (V$_{\textnormal{A}}$ and V$_{\textnormal{S}}$), which reflect the contributions from out-of-plane and in-plane SOTs, respectively. Since the magnetization of Py is in the plane,  $\uptau_{\textnormal{A}}$ and $\uptau_{\textnormal{B}}$ will be out-of-plane and  $\uptau_{\textnormal{S}}$ and $\uptau_{\textnormal{T}}$ will be in-plane. Conventional $\uptau_{\textnormal{A}}$ and $\uptau_{\textnormal{S}}$ contributions to the angular dependence of the V$_{\textnormal{A}}$ and V$_{\textnormal{S}}$ can be fitted as a function of cos($\varphi$)sin($2\varphi$) where $\varphi$ is the angle between magnetization and current (Fig. \ref{fig:wte2} (a)). Note that the cos($\varphi$) term comes from SOT effects and the sin($2\varphi$) term is from the derivative of anisotropic magnetoresistance since all SOTs modulate it in the ST-FMR signals. Unconventional SOTs originate from an out-of-plane spin polarization that is independent of $\varphi$. Therefore, V$_{\textnormal{B}}$ and V$_{\textnormal{T}}$ have a sin($2\varphi$) dependence. In summary, SOT contributions to V$_{\textnormal{A}}$ and V$_{\textnormal{S}}$ can be fitted to the first order using 
\begin{equation*}
\begin{split}
    V_{\textnormal{A}} (\varphi) = A\textnormal{cos}(\varphi)\textnormal{sin}(2\varphi)+B\textnormal{sin}(2\varphi)\\
    V_{\textnormal{S}} (\varphi) = S\textnormal{cos}(\varphi)\textnormal{sin}(2\varphi)+T\textnormal{sin}(2\varphi).
\end{split}
\end{equation*}
In a Pt/Py bilayer, only $A$ and $S$ are observed (see Fig. \ref{fig:wte2} (a)), which indicates finite $\uptau_{\textnormal{A}}$ and $\uptau_{\textnormal{S}}$. In the case of a WTe$_2$/Py device, the angular dependence of V$_{\textnormal{A}}$, as shown by the lower panel of Fig. \ref{fig:wte2} (b), apparently deviates the cos($\varphi$)sin$(2\varphi)$ function when the current is applied along $a$ axis. Unconventional $\uptau_{\textnormal{B}}$ is identified. It is worth noting that the out-of-plane $\uptau_{\textnormal{B}}$ was not observed when applying current along $b$ as shown in Fig. \ref{fig:wte2} (c), which agrees well with the symmetry consideration as explained in the previous section. The full angle dependence shows the the current-induced unconventional $\sigma_{\textnormal{B}}$ reduces when the current gradually changes the direction from the a-axis to the b-axis (Fig. \ref{fig:wte2} (d)). In contrast, the conventional $\sigma_{\textnormal{S}}$ does not show a clear change. Similar angle dependence of symmetry breaking-induced unconventional SOT was also reported in an artificial wedge structure.\cite{yu_switching_2014} \\

\begin{figure}
    \centering
    \includegraphics[width=16cm]{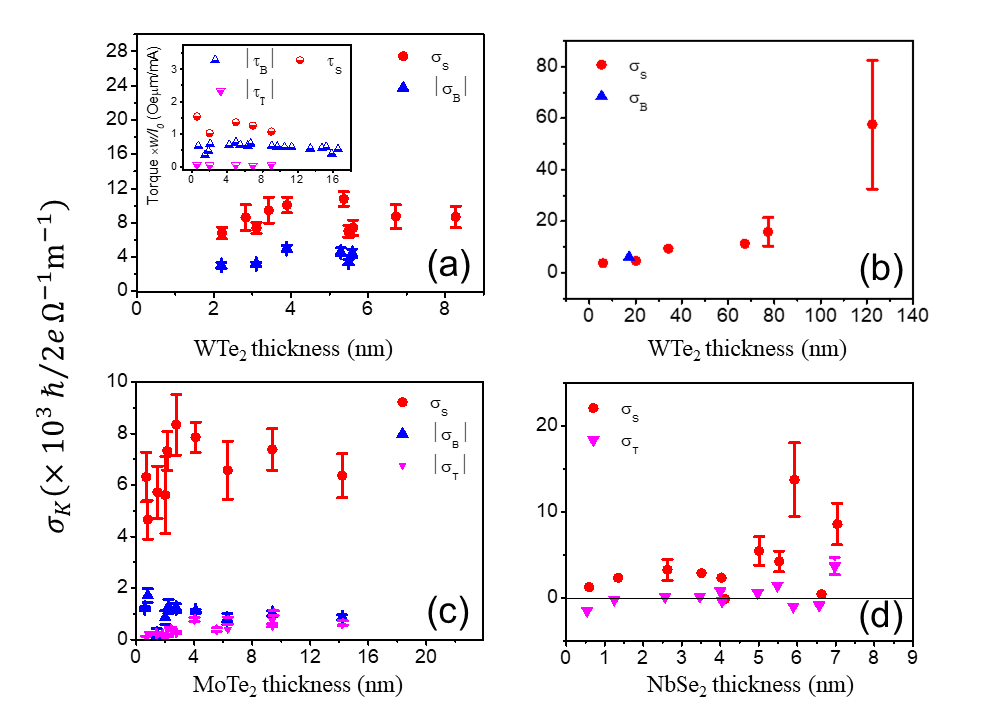}
    \caption{The thickness dependent of SOT or spin conductivities of various TMD/Py devices,  where $\sigma_{\textnormal{S}}$ stands for the conventional damping-like torque, $\sigma_{\textnormal{B}}$ stands for the out-of-plane damping-like torque and $\sigma_{\textnormal{T}}$ stands for the out-of-plane field-like torque. The spin conductivities as a function of the thickness of (a) and (b) WTe$_2$, (c) MoTe$_2$ and (d) NbSe$_2$. These figures are reproduced from Macneill $et$ $al$.,\cite{macneill_thickness_2017,macneill_control_2017} Shi $et$ $al$., \cite{shi_all-electric_2019} Stiehl $et$ $al$., \cite{stiehl_layer-dependent_2019} and Guimaraes $et$ $al$.. \cite{guimaraes_spin_2018}}
    \label{fig:tmdthick}
\end{figure}

The layer dependence of SOT in WTe$_2$ can be partially explained using the symmetry arguments. The magnitude of out-of-plane DL-SOT ($\uptau_{\textnormal{B}}$) shows no significant dependence on the WTe$_2$ thickness indicating the interfacial nature of $\uptau_{\textnormal{B}}$ as shown in Fig. \ref{fig:tmdthick} (a). However, the sign of the $\uptau_{\textnormal{B}}$ in the WTe$_2$/Py can be positive or negative which is in sharp contrast to the fixed sign in the HM/FM.\cite{macneill_control_2017} The authors attributed the sign change to the 2-fold screw symmetry of the bulk WTe$_2$. As shown in Fig. \ref{fig:wte2} (f), adjacent WTe$_2$ layers are related by a 180$^\circ$ rotation around the $c$ axis followed by a half-unit cell (one 2D layer) translation along the $c$ axis. Under the 180$^\circ$ rotation, current along the $a$ axis is reversed and the pseudovector $\uptau_{\textnormal{B}}$ remains unchanged. Therefore, neighboring layers could contribute to an opposite sign of $\uptau_{\textnormal{B}}$ and the overall $\uptau_{\textnormal{B}}$=0. In a WTe$_2$/Py bilayer, the 2-fold screw symmetry is broken and thus at the interface, finite $\uptau_{\textnormal{B}}$ can be expected. This argument was later directly proved by their subsequent work \cite{macneill_thickness_2017} in which SOTs of devices with a monolayer step have been measured by SHH measurements, as illustrated by Fig. \ref{fig:wte2} (f). The existence of $\uptau_{\textnormal{B}}$ is evidenced by the $\textnormal{cos}(2\varphi)$ angle dependence in the SHH voltage as shown in Fig. \ref{fig:wte2} (g). $\uptau_{\textnormal{B}}$ of two Hall bar regions across a monolayer step have opposite signs. As shown in Fig. \ref{fig:wte2} (g), the signs of $\uptau_{\textnormal{B}}$ are opposite for a monolayer and a bilayer WTe$_2$. In contrast, in all devices without a monolayer step, two Hall bar regions have the same sign. In the WTe$_2$/Py, $\uptau_{\textnormal{A}}$ shows a significant dependence on the layer thickness, which is dominated by the current-induced Oersted field.\cite{macneill_thickness_2017} Similar to $\uptau_{\textnormal{B}}$, the thickness independence of $\uptau_{\textnormal{S}}$ of the WTe$_2$/Py device also indicates an interfacial origin. \cite{macneill_thickness_2017,macneill_control_2017} Interestingly, at very large thickness, Shi $et$ $al$. \cite{shi_all-electric_2019} showed the $\uptau_{\textnormal{S}}$ is much enhanced (see Fig. \ref{fig:tmdthick} (b)).  \\  
\begin{table*}
	\caption{\label{tab_SOTexperiment} The crystal structures and spin conductivities of 2D/FM heterostructures}
	\resizebox{\textwidth}{!}
	{
		\begin{tabular}{lllll}
			\hline \hline
			\multirow{2}{*}{Authors}&\multirow{2}{*}{Materials}&Point group&Spin conductivity&\multirow{2}{*}{Note}\\
			&&(space group)&($\times$ 10$^3$ ($\hbar/2e$)($\Omega m)^{-1}$)&\\ \hline
			\multirow{2}{*}{Shao $et$ $al$. \cite{shao_strong_2016}}&MoS$_2$/CoFeB&P6/mmc&$\sigma_{\textnormal{A}}$ = 2.9&\\
			&WSe$_2$/CoFeB&P6/mmc&$\sigma_{\textnormal{A}}$ = 5.5&\\\hline
			\multirow{2}{*}{MacNeil $et$ $al$. \cite{macneill_control_2017}}&\multirow{3}{*}{WTe$_2$/Py}&\multirow{3}{*}{Pmn2$_1$}
			&$\sigma_{\textnormal{A}}$ = 9 $\pm$ 3&\multirow{3}{*}{thickness independent $<$ 10 nm}\\
			&&&$\sigma_{\textnormal{S}}$ = 8 $\pm$ 2&\\
			&&&$\sigma_{\textnormal{B}}$ = 3.68 $\pm$ 0.8&\\
			\multirow{2}{*}{Shi $et$ $al$. \cite{shi_all-electric_2019}}&\multirow{3}{*}{WTe$_2$/Py}&\multirow{3}{*}{ Pmn2$_1$}&$\sigma_{\textnormal{A}}$ due to Oersted field&\\
			&&&$\sigma_{\textnormal{S}}$ = 4-60& thickness dependent\\
			&&&$\sigma_{\textnormal{B}}$ = 6& \\\hline
			\multirow{4}{*}{Stiehl $et$ $al$. \cite{stiehl_layer-dependent_2019}}&\multirow{4}{*}{MoTe$_2$/Py}&\multirow{4}{*}{P2$_1$m}
			&$\sigma_{\textnormal{A}}$ due to Oersted field&\multirow{4}{*}{thickness dependent}\\
			&&&$\sigma_{\textnormal{S}}$ = 4.7-8.2& \\
			&&&$\sigma_{\textnormal{B}}$ = 0-1.8&\\
			&&&$\sigma_{\textnormal{T}}$ = 0-1.0&\\\hline
			\multirow{3}{*}{Guimaraes $et$ $al$. \cite{guimaraes_spinorbit_2018}}&\multirow{3}{*}{NbSe$_2$/Py}&\multirow{3}{*}{P6$_3$/mmc}
			&$\sigma_{\textnormal{A}}$ due to Oersted field&thickness dependent\\
			&&&$\sigma_{\textnormal{S}}$ = 0-13&and random $\sigma_{\textnormal{T}}$ due to\\
			&&&$\sigma_{\textnormal{T}}$ = -2-3.5&uncontrollable strain effect\\\hline
			Xu $et$ $al$.\cite{xu_high_2020}&PtTe$_2$/Py&P$\bar{3}$m1&20-160&thickness dependent\\
			\hline\hline
		\end{tabular}
	}
\end{table*} 
Monoclinic 2D materials also fulfill the symmetry requirement for generating the out-of-plane $\uptau_{\textnormal{DL}}$ despite the fact that monoclinc 2D materials are inversion symmetric by themselves. Stiehl $et$ $al$.\cite{stiehl_layer-dependent_2019} have investigated the SOTs of the $\beta$ or 1T' phase MoTe$_2$. Its lattice structure belongs to the C$_{2h}$ point group (space group: P2$_1$/m) which has a 2-fold screw axis along $b$ axis and a mirror plane perpendicular to it. When interfacing to an FM layer, the out-of-plane $\uptau_{\textnormal{DL}}$ is allowed by symmetry when sending current along the $b$ axis. The SOTs were measured by the ST-FMR technique and the observed SOTs in MoTe$_2$/Py exactly follow this symmetry consideration. Conventional SOTs and the out-of-plane $\uptau_{\textnormal{DL}}$ are all observed when the applied current is along $b$ axis. \cite{stiehl_layer-dependent_2019} Similar to WTe$_2$, current-induced unconventional $\sigma_{\textnormal{B}}$ reduces when the current gradually changes the direction from the a-axis to the b-axis (Fig. \ref{fig:wte2} (e)). The layer dependence of $\sigma_{\textnormal{B}}$ and $\sigma_{\textnormal{S}}$ reveals an interfacial origin (Fig. \ref{fig:tmdthick} (c)). \\
TaTe$_2$ has a space group of C$2/m$. The layer stacking direction is along its $b$ axis as indicated by Fig. \ref{fig:my_latticestructure} (k). According to its space group notation, there is a 2-fold rotation axis about $b$ axis and a mirror plane perpendicular to it. The surface normal direction deviates from the $b$ axis, hence the 2-fold rotation symmetry is broken. Unconventional SOTs are allowed by symmetry in this material. Stiehl $et$ $al$.\cite{stiehl_current-induced_2019} have studied SOTs in TaTe$_2$/Py. However, no unconventional SOTs were observed, which could be due to the weak SOC in TaTe$_2$. \cite{stiehl_current-induced_2019} In the same work, the authors examined the importance of considering the resistance asymmetry of low-symmetry 2D materials in determining the SOTs. \\
Some results indicate a possibility to generate the unconventional SOTs in highly symmetric TMDs by using strain effect. Guimaraes $et$ $al$. \cite{guimaraes_spinorbit_2018} have reported SOTs in NbSe$_2$/Py as shown in Fig. \ref{fig:tmdthick} (d). NbSe$_2$ has a hexagonal structure belonging to a space group of P6$_3$/mmc which preserves more than one mirror plane perpendicular to an in-plane axis. Like 2H-MoS$_2$, unconventional SOTs are prohibited by symmetry requirements. However, an unconventional FL-SOT ($\uptau_{\textnormal{T}}$) was observed. The magnitude and sign of the ($\uptau_{\textnormal{T}}$) don't show any trends with the NbSe$_2$ thickness and they vary strongly among samples. The authors attributed the arising torque to a strain effect induced by the device fabrication which cannot be controlled at this stage. This strain effect breaks the rotation symmetry and reduces the number of mirror planes to one. As a result, unconventional SOTs are allowed. The crystal structure and spin conductivity of TMDs are summarized in Table \ref{tab_SOTexperiment}.\\
At last, we would like to discuss the origin of the unconventional SOTs. There are two possible mechanisms for the unconventional SOTs. The first one is SHE. This mechanism is recently evidenced by Song $et$ $al$.\cite{song_coexistence_2020} and Safeer $et$ $al$. \cite{safeer_large_2019} that an out-of-plane $\sigma$ attributing to the intrinsic SHE and inverse SHE has been observed in a few layers MoTe$_2$ with a non-local spin detection. The second mechanism could attribute to the Rashba effect that originates from the hybridization between the FM layer and the top layers.\cite{macneill_control_2017} However, there are still phenomena that cannot be explained using these two models. For example, the layer dependence of $\uptau_{\textnormal{B}}$ in MoTe2/Py shows that $\uptau_{\textnormal{B}}$ is observed in mono- and triple-layered MoTe$_2$ but not in double-layered MoTe$_2$\cite{stiehl_layer-dependent_2019} (Fig. \ref{fig:tmdthick} (c)). The exact mechanism for the absence of $\uptau_{\textnormal{B}}$ in the bilayer MoTe$_2$ remains unclear. \\

\subsection{Growth of 2D Materials for Spintronic Applications}
\begin{figure}
    \centering
    \includegraphics[width=16cm]{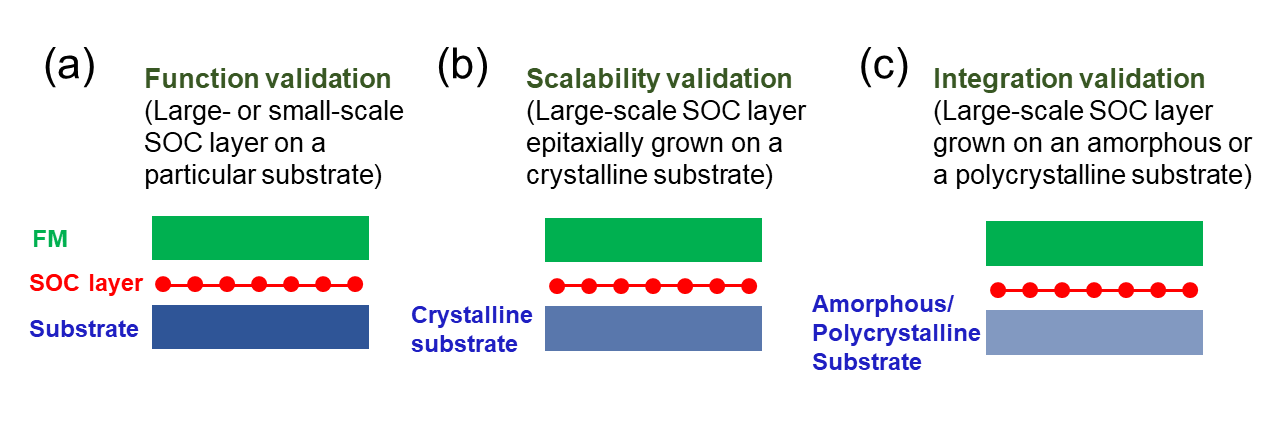}
    \caption{Steps toward practical SOT devices based on crystalline SOC layers, including 2D materials.}
    \label{fig:growth}
\end{figure}
Up to now, most of 2D materials used for SOT studies are not scalable due to the small sample size or difficulty in integration with the CMOS technology. For practical SOT device applications, it is critical to validate three important concepts (see Fig. \ref{fig:growth}). First, it is function validation, where a 2D material grown on a particular substrate or exfoliated on a substrate is used. As described above, large-scale TIs have been grown using molecular beam epitaxy (MBE) on different substrates\cite{kou_epitaxial_2019} and they have been used to demonstrate a giant SOT efficiency. In contrast, TMDs are exfoliated on a substrate, such as a silicon wafer with thermal oxide, and then FM layers are deposited on them for SOT studies. In this step, two important things need to be validated. First, the SOT magnitudes need to precisely quantified. Second, the ratio of SOT switching current to the thermal stability factor needs to be assessed. It would be much better if the SOT switching current is determined when the FM layer is in the single domain regime, which is usually achieved by nanopatterning the FM layer into nanodot. So far, most of SOT studies on 2D materials are at this stage. \\
Second, it is scalability validation. While 2D materials prepared by mechanical exfoliation could have exceptional crystalline quality, they are not scalable. There are four commonly used methods to grow crystalline 2D materials: MBE, chemical vapor deposition (CVD), metal-organic CVD (MOCVD), and magnetron sputtering. For a review of the MBE growth of TIs, readers can check Kou $et$ $al$..\cite{kou_epitaxial_2019} For a review of the TMD growth using the first three approaches, readers can check Manzeli $et$ $al$..\cite{manzeli_2d_2017} Early attempts to deposit high-quality TMD materials using magnetron sputtering have been reported by Huang $et$ $al$.\cite{huang_large-area_2017}, Samassekou $et$ $al$. \cite{samassekou_viable_2017} and Wang $et$ $al$., \cite{wang_high_2018} where only electrical and optical properties were measured. Recently, Xu $et$ $al$.\cite{xu_high_2020} used a two-step process to grow a large-scale PtTe$_2$ and examined its spintronic properties, revealing a very high $\sigma_{\textnormal{S}}$. In the future, spintronic properties of these 2D materials over a large scale need to be systematically quantified. Note that it is critical to ensure single crystallinity and single thickness of the 2D material over the wafer scale. The reason is twofold. First, for TIs, the thickness affects the SOT magnitude significantly\cite{wang_room_2017} and thus thickness fluctuations may cause variation in switching current. Second, to utilize the out-of-plane damping-like SOT generated by TMDs, random crystalline domains may contribute random out-of-plane FL-SOT directions, which make large-scale production of SOT devices impractical. Therefore, it is necessary to use a single-crystalline substrate (or buffer).\\
Third, it is integration validation. Eventually, SOT devices need to be integrated with standard CMOS technology using the back-end-of-line (BEOL) process, where the crystalline SOC layer needs to interface an amorphous or polycrystalline substrate such as silicon oxide or metals. Even with a buffer layer, the direct growth of single-crystalline SOC layer could be very challenging on an amorphous or polycrystalline substrate over a large scale. Alternatively, transferring a large-scale crystalline SOC layer could be beneficial. In this approach, one needs to grow a high-quality 2D material on a crystalline material or buffer and then transfer it on to the desired substrate. Readers can check Lin $et$ $al$.\cite{lin_two-dimensional_2019} for recent progress.

\section{Conclusion and Outlook}
\begin{figure}
    \centering
    \includegraphics[width=16cm]{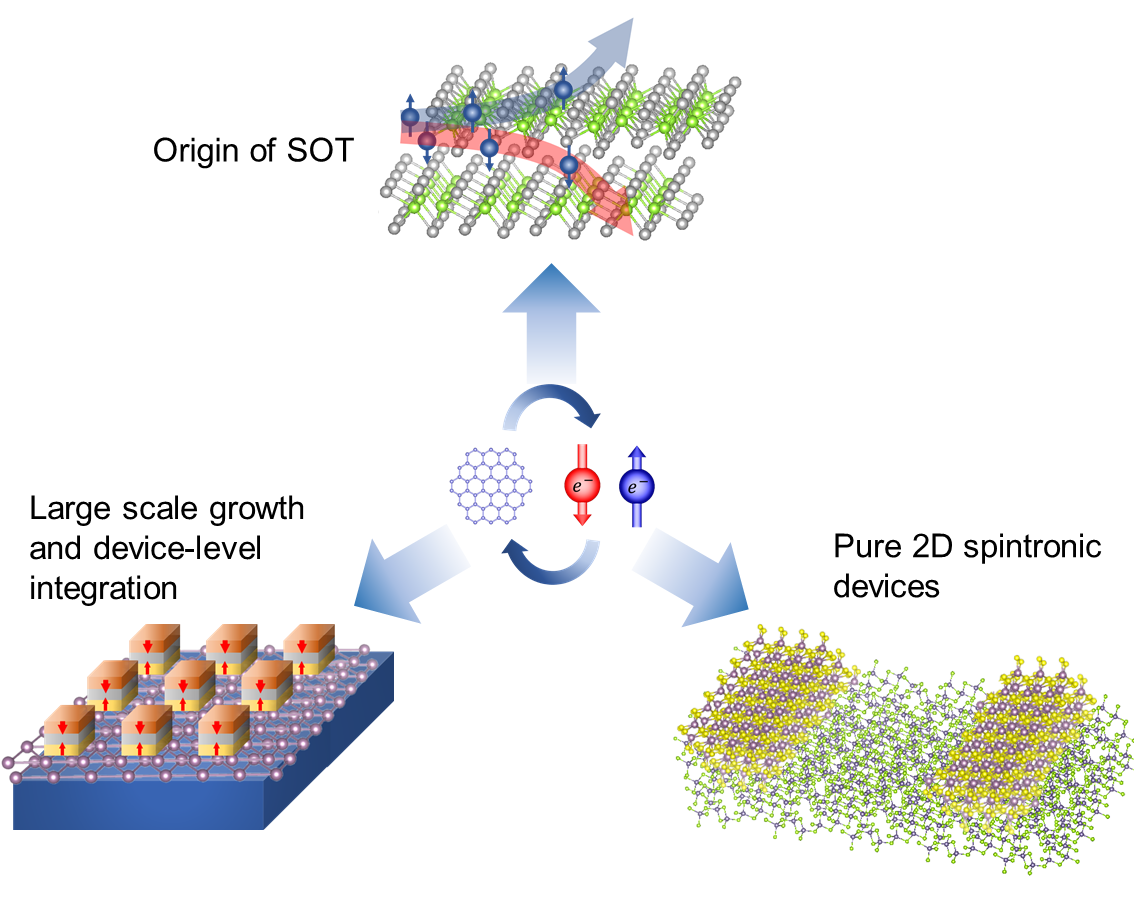}
    \caption{Research directions of 2D material based SOT and spintronic devices.}
    \label{fig:outlook}
\end{figure}
In this review, we have presented the systematic analysis of SOTs based on the symmetry argument and show how this method can be used to predict and analyze the SOTs in 2D materials with various crystal symmetries. We highlight the recent progress of giant SOTs in TI-based heterostructures and unconventional out-of-plane damping-like SOTs in TMD-based heterostructures that could be very beneficial for energy-efficient SOT devices. Based on this review, we suggest the following three important research directions in this field as illustrated in Fig. \ref{fig:outlook}.\\
\textbf{(a) understanding the origin of SOTs.} While the symmetry argument could provide a qualitative prediction of SOTs, a quantitative prediction relies on the understanding of microscopic picture of SOT generation. Investigating other spin-charge interconversion phenomena such as spin pumping,\cite{mendes_efficient_2018} spin injection,\cite{song_coexistence_2020,safeer_large_2019} spin thermoelectric effect,\cite{dau_valley_2019} and electrically induced magnetization (magnetoelectricity)\cite{lee_valley_2017} in 2D material-based heterostructures would also be helpful since they are highly correlated to the SOT physics. 2D materials are layered systems, which are highly anisotropic in crystal structures. The fundamental studies of spin generation and transport in out-of-plane direction are still very limited. As a result, the thickness dependence of SOTs in 2D materials are not well understood.\cite{stiehl_layer-dependent_2019,shi_all-electric_2019}\\
\textbf{(b) purely 2D material-based spintronic devices.} Recent discoveries of magnetic 2D materials, such as CrI$_3$,\cite{huang_layer-dependent_2017} Cr$_2$Ge$_2$Te$_6$\cite{gong_discovery_2017} and Fe$_3$GeTe$_2$,\cite{deng_gate-tunable_2018} could enable the integration of non-magnetic 2D materials and magnetic 2D materials for spintronic applications.\cite{gong_two-dimensional_2019,zhang_van_2019} On the one hand, magnetic 2D materials possess desirable properties for spintronic applications. The excellent gate tunability of magnetic 2D materials allows for exploring magnetoelectric phenomena and applications.\cite{deng_gate-tunable_2018} Meanwhile, layered antiferromagnetic 2D materials like CrI$_3$ exhibit a giant spin-filter tunnel magnetoresistance effect.\cite{song_giant_2018,klein_probing_2018} Furthermore, magnetic ordering of magnetic 2D materials can be manipulated by SOTs.\cite{alghamdi_highly_2019,wang_current-driven_2019} On the other hand, non-magnetic 2D materials could be an excellent spin source owing to their large and multidirectional SOTs as we explained in this review. Besides practical applications, the integration could allow us to explore the exotic phases and interactions in 2D magnetic heterostructures,\cite{zhao_magnetic_2020} which are of fundamental interests.  \\
\textbf{(c) pushing 2D materials towards practical SOT devices.} For practical applications, two important factors need to be taken into account. First, 2D materials need to be integrated into CMOS-compatible magnetic materials with PMA.\cite{shao_room_2018} Currently, CoFeB/MgO-based MRAM technologies have been mature\cite{garello_sot-mram_2018,golonzka_mram_2018,song_demonstration_2018} and it would be very beneficial if 2D materials could be integrated with them. Second, one needs to develop methods of wafer-scale integration of 2D materials and CMOS-compatible substrates. Potential approaches have been discussed in the large-scale growth section.

\begin{acknowledgement}
We thank Wen-Yu He and K. T. Law for insightful discussions. The authors are supported by HKUST ECE start-up fund. The authors use VESTA and $Crystallography Open Databases$ for 3D visualizing of crystals.

\end{acknowledgement}

\bibliography{references.bib}

\end{document}